\documentclass{l4dc2024}

\usepackage{caption}

\usepackage{adjustbox}
\usepackage{array}
\usepackage{wrapfig}

\newcolumntype{R}[2]{%
    >{\adjustbox{angle=#1,lap=\width-(#2)}\bgroup}%
    l%
    <{\egroup}%
}

\newtheorem{problem}{Problem}

\title[Data-Driven Modeling and Verification of Perception-Based Autonomous Systems]{Data-Driven Modeling and Verification of Perception-Based Autonomous Systems}
\usepackage{times}

\author{%
 \Name{Thomas Waite} \Email{waitet@rpi.edu}\\
 \addr Dept. of Computer Science, Rensselaer Polytechnic Institute
 \AND
 \Name{Alexander Robey} \Email{arobey1@seas.upenn.edu}\\
 \addr Dept. of Electrical and Systems Engineering, University of Pennsylvania
 \AND
 \Name{Hamed Hassani} \Email{hassani@seas.upenn.edu}\\
 \addr Dept. of Electrical and Systems Engineering, University of Pennsylvania
 \AND
 \Name{{George J.} {Pappas}} \Email{pappasg@seas.upenn.edu}\\
 \addr Dept. of Electrical and Systems Engineering, University of Pennsylvania
 \AND
 \Name{Radoslav Ivanov} \Email{ivanor@rpi.edu}\\
 \addr Dept. of Computer Science, Rensselaer Polytechnic Institute
}

\begin{document}

\maketitle

\begin{abstract}
This paper addresses the problem of data-driven modeling and verification of perception-based autonomous systems. We assume the perception model can be decomposed into a canonical model (obtained from first principles or a simulator) and a noise model that contains the measurement noise introduced by the real environment. We focus on two types of noise, benign and adversarial noise, and develop a data-driven model for each type using generative models and classifiers, respectively. We show that the trained models perform well according to a variety of evaluation metrics based on downstream tasks such as state estimation and control. Finally, we verify the safety of two systems with high-dimensional data-driven models, namely an image-based version of mountain car (a reinforcement learning benchmark) as well as the F1/10 car, which uses LiDAR measurements to navigate a racing track.
\end{abstract}

\section{Introduction}
\label{sec:intro}

From self-driving cars (\cite{waymo}) to taxi helicopters (\cite{volocity}), the last few years have seen the development of a number of impressive autonomous systems. As the complexity of these systems increases, however, so does the concern for their safety. In fact, we have already witnessed accidents involving systems as diverse as autonomous cars (e.g.,~see the reports by the \cite{tesla_report} and the~\cite{uber_report}), chess-playing robots (as reported by \cite{chess_robot}) and autonomous aircraft (analyzed in an \cite{boeing_report} report). Furthermore, the United States government recently recorded 367 crashes involving autonomous cars over a 10-month period (\cite{nhtsa}). In order to prevent such incidents, it is essential that we verify the safety of autonomous systems before they are deployed in the wild.

Unlike classical control systems, modern autonomous systems introduce an extra layer of complexity since they operate in complex environments, which are perceived through high-dimensional measurements such as LiDAR scans and camera images. In turn, these measurements are processed by neural networks (NNs) used for estimation and control. Verifying such a closed-loop system at design-time poses two significant challenges: 1) environment models are difficult to develop from first principles due to unexpected noise, e.g., reflected LiDAR rays (\cite{ivanov20a}); 2) NNs are not robust to even small input perturbations (\cite{szegedy13}) and distribution shifts (\cite{recht19}), which may reduce the utility of verification performed against an imperfect model.

To overcome these challenges, in this paper we propose a compositional verification approach that uses data-driven environment models. We compose two types of models: 1) a canonical environment model (e.g., during daytime with perfect visibility) that is obtained from first principles or through a simulator; 2) a data-driven noise model that is trained on real data to augment the canonical model with data artifacts observed during real system operation (e.g., blurred images or reflected LiDAR rays). This approach is motivated by recent work by \cite{katz22} on developing (and verifying) canonical models from real observations as well as work by \cite{wu23} on using generative models for certifying the robustness of NNs to real-world distribution shifts. The compositional model has several benefits over a single monolithic model: 1)~the canonical model need not be data-driven, thereby reducing the data requirements for training the perception model; 2)~different noise models can be composed with the canonical model to capture diverse scenarios; 3)~individual noise models can be smaller, thereby alleviating verification scalability challenges.

Training noise models presents an interesting challenge since some types of noise are easier to capture using generative models than other types. For example, continuous noise (e.g., blur or contrast) is fairly benign and is easily learned by a generative model. However, discontinuous noise (e.g., reflected LiDAR rays or lens flare) is more adversarial and cannot be perfectly learned using a continuous model. We handle these two cases separately: 1)~for benign noise, we use generative models, e.g.,~variational autoencoders as introduced by \cite{kingma13}; 2)~for adversarial noise (defined as high-frequency large deviation from the canonical model), we train classifiers that indicate which part of the canonical measurement is affected and can be replaced with the effect of the adversarial noise. In fact, our experiments suggest that it is exactly the discontinuous noise that causes the largest deviations in control performance as compared to the canonical environment.

To evaluate the compositional data-driven modeling approach, we present two verification case studies: 1) Mountain Car (MC), which is a reinforcement learning (RL) benchmark available in \cite{mc}; 2)~the~\cite{f1tenth}, which is a 1/10-scale autonomous racing car platform. In MC, we train an image-based control pipeline and verify that the car reaches the goal for a range of benign noises, including blur and contrast. In the F1/10 case, we use existing LiDAR traces (\cite{ivanov20a}) to train an adversarial noise classifier. The noise model's quality is evaluated on the downstream control task, by comparing the control output on real vs. modeled LiDAR scans. Finally, we verify that the F1/10 car, coupled with a robust denoiser, can safely navigate an environment under adversarial LiDAR noise that was shown to be correlated with crashes by \cite{ivanov20a}. In both cases, we performed the verification using our tool, Verisig (\cite{ivanov19,ivanov20,ivanov2021b}).

In summary, the contributions of this paper are as follows: 1) a compositional data-driven method for developing perception models used in verification; 2) a noise-specific approach that uses generative models for benign noise and classifiers for adversarial noise; 3) two verification case studies illustrating the benefit of data-driven modeling and verification.

\vspace{-5px}
\paragraph{Related work.} Verification of standard (non-neural) control systems is a mature field. A classical approach to reachability is by using Hamilton-Jacobi methods (\cite{mitchell02,chen18}). Another class of techniques approximate reachable sets using Taylor models (\cite{makino03,chen12}, ellipsoids (\cite{althoff15}), and polytopes (\cite{chutinan03}). Finally, there also exist methods that cast the problem as a satisfiability modulo theory program (\cite{gao13,kong15}). More recently, a number of works were developed for open-loop verification of NNs, such as input-output robustness, e.g., works by \cite{dutta18,ehlers17,fazlyab19,gehr18,katz17,wang18,weng18,tran20,wang21}. Methods also exist for verification of closed-loop systems with NN components such as those by \cite{ivanov19,huang19,dutta19,sun19,tran19,bogomolov19,dreossi19}.

For systems with high-dimensional perception, \cite{katz22} developed a verification method by training a generative image model using simulated data. \cite{ivanov20a} used a first-principles LiDAR model to verify the safety of the F1/10 car as it navigates a racing track. \cite{hsieh22} verified perception-based systems using approximate abstractions of the perception system. Although these works are good first steps, they do not account for the distribution shift between the canonical model and the real data that would be captured by our noise model. Approaches also exist (\cite{dawson22,robey20,robey21b}) for learning safe barrier certificates from images but these do not provide worst-case guarantees on the learned certificate. Finally, a number of works exist (\cite{hanspal23,robey2020model,wu23}) that use model-based generative NNs for open-loop verification and robustness purposes; we borrow ideas from these approaches (e.g., model-based training) when tackling the closed-loop problem.
\begin{figure}[t]
  \centering 
  \includegraphics[width=0.7\linewidth]{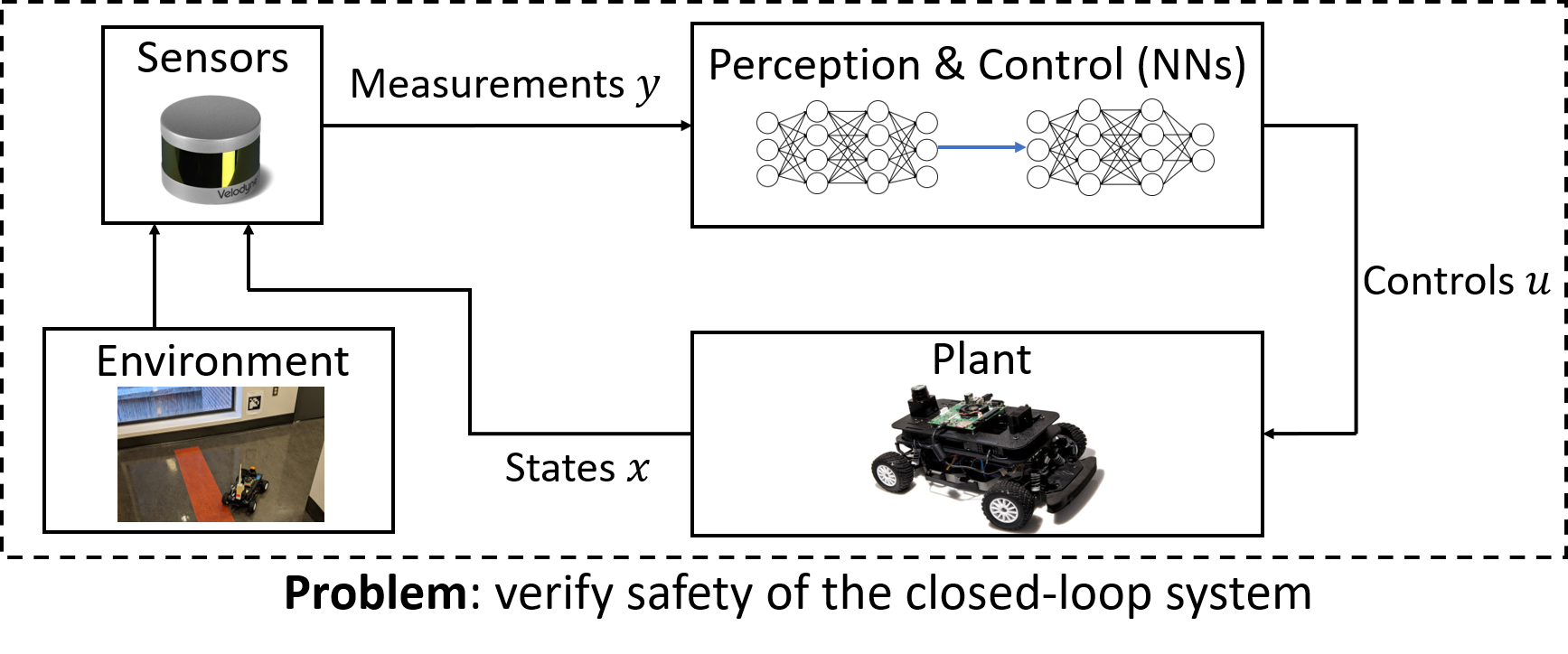}
  \vspace{-10px}
  \caption{Overview of the problem considered in this paper.}
  \label{fig:overview} \vspace{-4mm}
\end{figure}
\vspace{-10px}
\section{Problem Formulation}
\label{sec:problem}

This section formalizes the problem addressed in this paper. We consider a closed-loop autonomous system, as illustrated in Figure~\ref{fig:overview}. Formally, we are given a dynamical system of the sort:\footnote{A continuous-time formulation can also be handled by the proposed framework.}
\vspace{-7px}
\begin{align}
\label{eq:system_model}
\begin{split}
x_{k+1} &= f(x_k, u_k)\\
y_k &= g(x_k) = g_n(x_k, \delta_k) \circ g_c(x_k)\\
u_k &= h(y_k),
\end{split}
\end{align}
\vspace{-13px}
$\\$
where $x \in \mathbb{R}^n$ is the system state (e.g., position, velocity), $y \in \mathbb{R}^m$ are the measurements (e.g., camera images and LiDAR scans) and $u \in \mathbb{R}^p$ are the controls. The dynamics $f$ are known. The observation model $g$ is unknown but we assume it can be represented as the composition of a canonical environment model, $g_c$, and a noiser, $g_n$. The canonical model is known (either developed from first principles or a simulator); the noiser is unknown as it contains the noise profile of the real environment the system operates in; the parameter $\delta_k$ specifies the noise intensity (e.g., blur level). Finally, $h$ encodes the perception/control pipeline and contains one or more NNs.

To train the noise model, we assume we are given a training set $\mathcal{D}=\{(x_i,y_i)\}$ of states and measurements. The training set is collected by running the system (e.g., manually) in the real environment. This setting is inspired by the one considered by \cite{dean20}, where $\mathcal{D}$ is used to obtain probably approximately correct (PAC) bounds on perception error. An important challenge is developing a metric for evaluating the noise model's quality; metrics defined on the high-dimensional measurement space may be misleading, as distances (e.g., pixel differences) may not be semantically meaningful. Thus, part of the problem is to develop a semantically meaningful metric to evaluate the noise model. We now state the two problems considered in this work.

\vspace{-4px}
\begin{problem}
Consider the closed-loop system in~\eqref{eq:system_model}. Given a training dataset, $\mathcal{D}=\{(x_i,y_i)\}$, train a noise model, $g_n$. Furthermore, develop a semantically meaningful evaluation metric for $g_n$.
\end{problem}
\vspace{-10px}
\begin{problem}
Consider the closed-loop system in~\eqref{eq:system_model} where $g_n$ is well trained. Given an initial set $\mathcal{X}_0$, the problem is to verify a safety property $\phi$ (e.g., no collisions) of the reachable states $x_k$:
\vspace{-5px}
\begin{equation}
\label{eq:cl_prop}
(x_0 \in \mathcal{X}_0) \Rightarrow \phi(x_k),\ \forall k \ge 0.
\end{equation}
\end{problem}

\begin{figure}[t]
  \centering
  \floatconts{fig:mc_examples}
{ \caption{Examples of benign noise in the MC environment.}}
{%
\centering
\subfigure[Canonical][b]{%
\label{fig:mc_canonical}
\includegraphics[width=.24\linewidth]{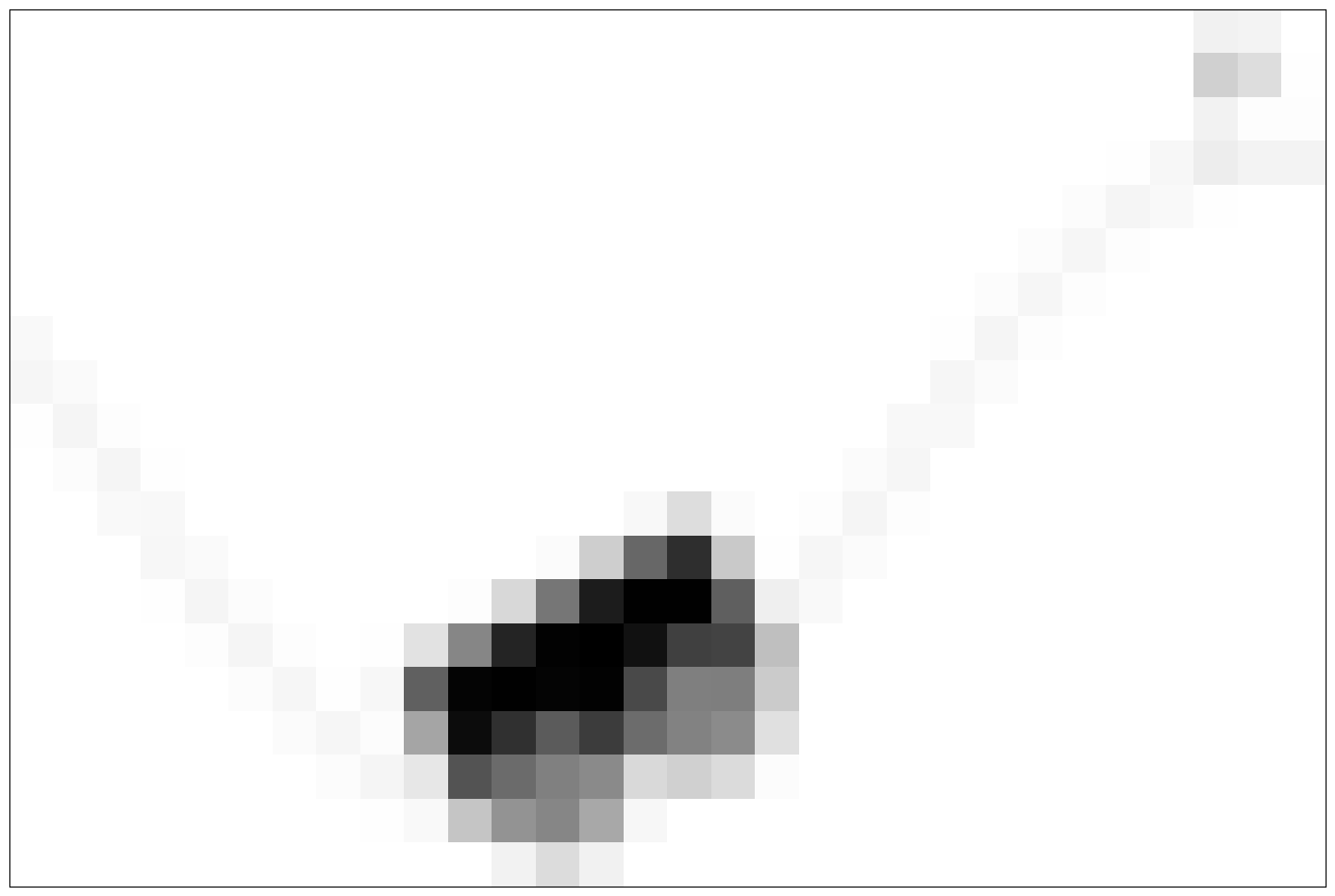}
}
\subfigure[Reduced Contrast][b]{%
\label{fig:mc_contrasted}
\includegraphics[width=.24\linewidth]{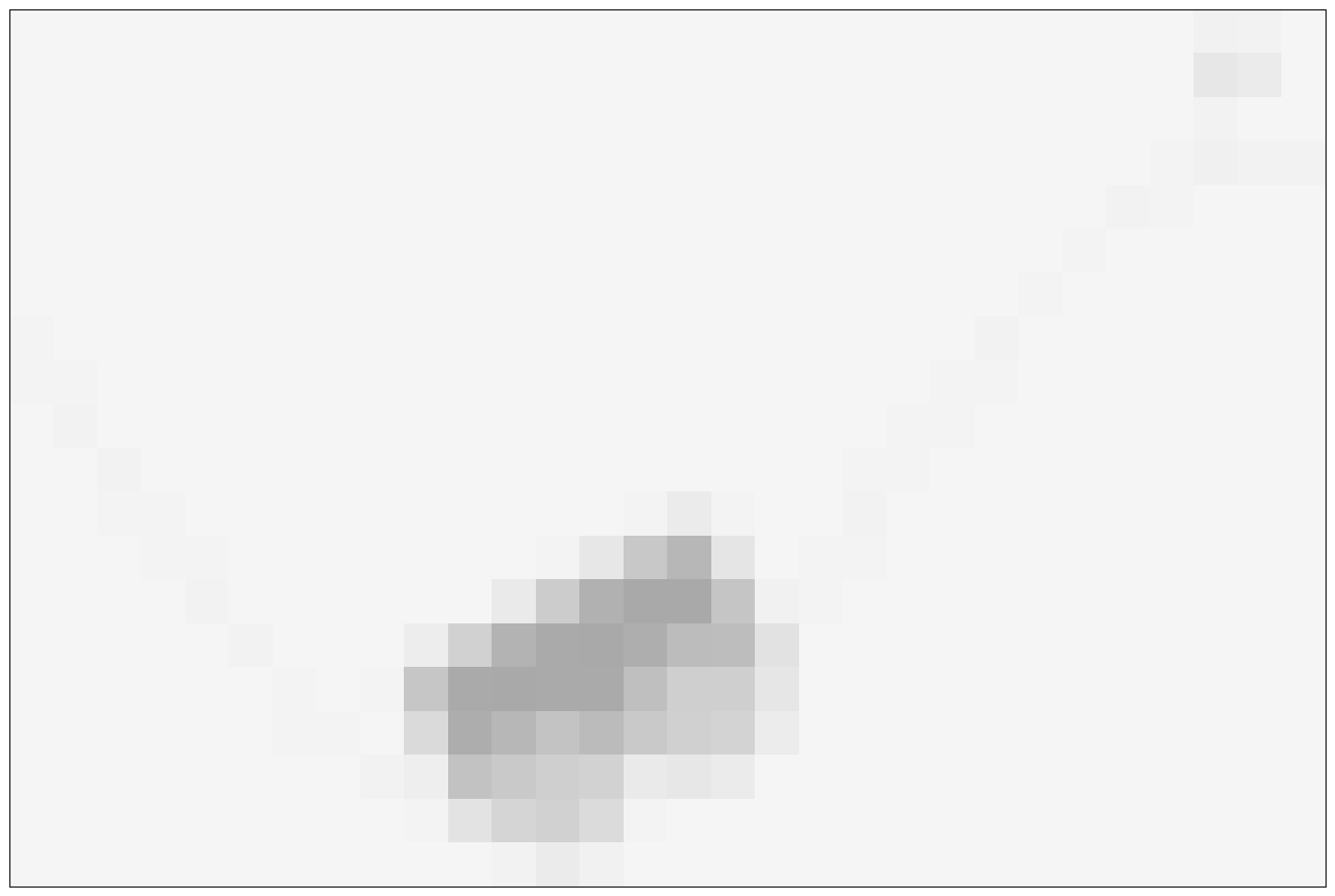}
}
\subfigure[Blur][b]{%
\label{fig:mc_blurred}
\includegraphics[width=.24\linewidth]{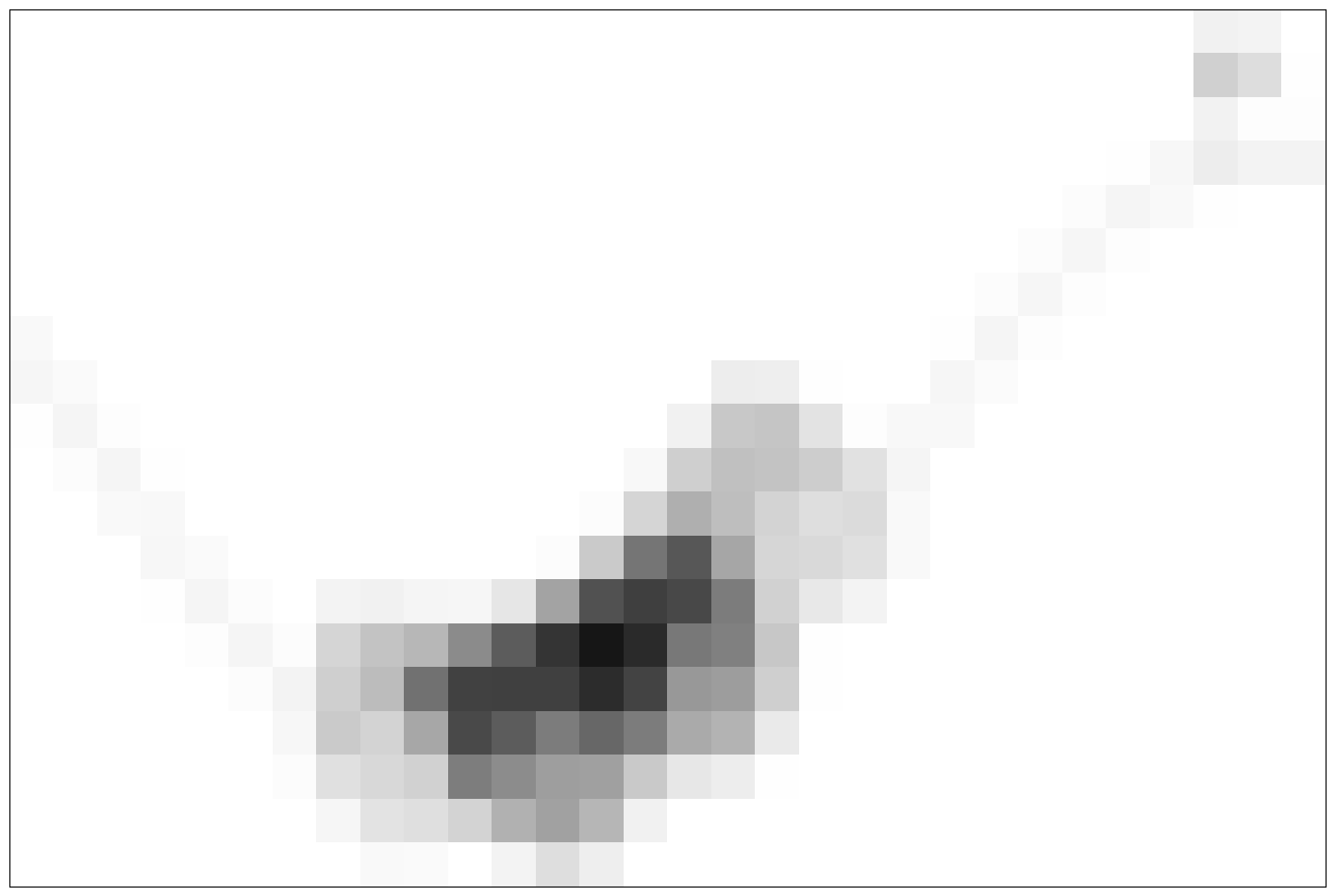}
}
\subfigure[Contrast + Blur][b]{%
\label{fig:mc_composite}
\includegraphics[width=.24\linewidth]{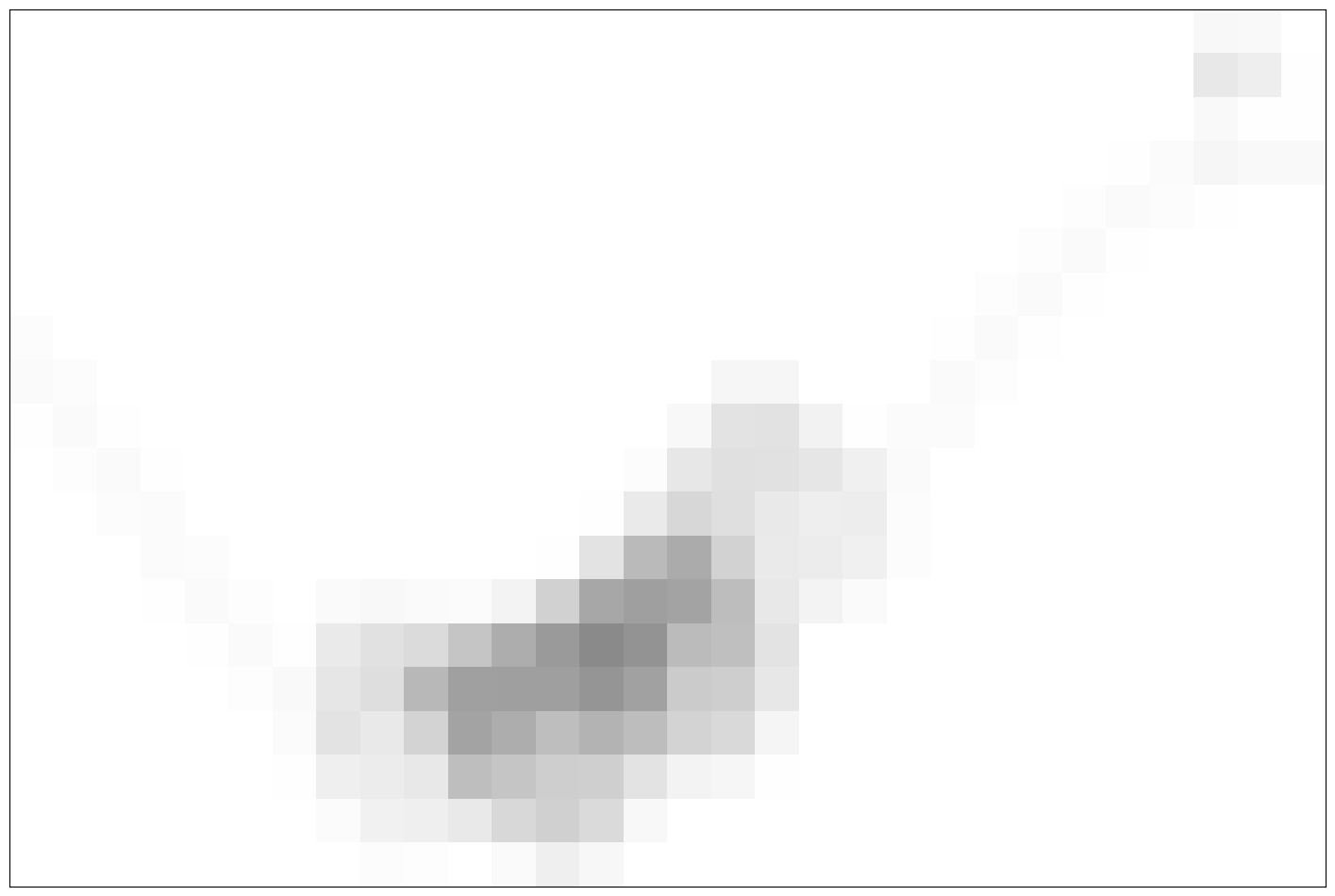}
} 
\vspace{-10mm}
}\vspace{-5mm}  
\end{figure}

\vspace{-25px}
\section{Motivating Examples}
\label{sec:examples}

This section presents two examples to illustrate different system setups, i.e., an image-based vs. a LiDAR-based system, as well as different noise types, namely benign vs. adversarial. Finally, while the MC example is purely synthetic, the F1/10 case study uses real data to train the noise model.

\vspace{-5px}
\subsection{Mountain Car: An Image-Based System Operating Under Benign Noise}
MC is an RL benchmark where the task is to drive an underpowered car up a hill, as illustrated in Figure~\ref{fig:mc_canonical} (we modified the environment by reducing the image size and making the car bigger). Since the car does not have enough power, it needs to learn to drive up the left hill so as to gather momentum and get to the goal on the right. MC consists of two states, position and velocity, which are both observed by the controller in the original MC environment. 
In this paper, we consider the case where the controller observes an image (such as the one in Figure~\ref{fig:mc_canonical}), coupled with velocity.\footnote{As it is impossible to infer velocity from a single image, we leave the case of using multiple images for future work.} 

Given an image-based controller, we are interested in verifying the system's robustness to environment noise. As illustrated in Figures~\ref{fig:mc_contrasted} and~\ref{fig:mc_blurred}, we consider two types of noise, contrast and blur (as well as their combination, shown in Figure~\ref{fig:mc_composite}), that aim to capture realistic operating conditions. Different contrast corresponds to changing lighting conditions, e.g., time of day, whereas blur occurs at higher driving speeds. Both noise types are continuous and are classified as benign. Section~\ref{sec:modeling} presents a method to train and verify a generative model for such types of noise.

\begin{figure}[t]
  \centering
  \floatconts{fig:f11_examples}
{ \caption{Illustration of the F1/10 car, including canonical and real scans featuring adversarial noise.}}
{%
\subfigure[LiDAR illustration][b]{%
\label{fig:f110_overview}
\includegraphics[width=.24\linewidth]{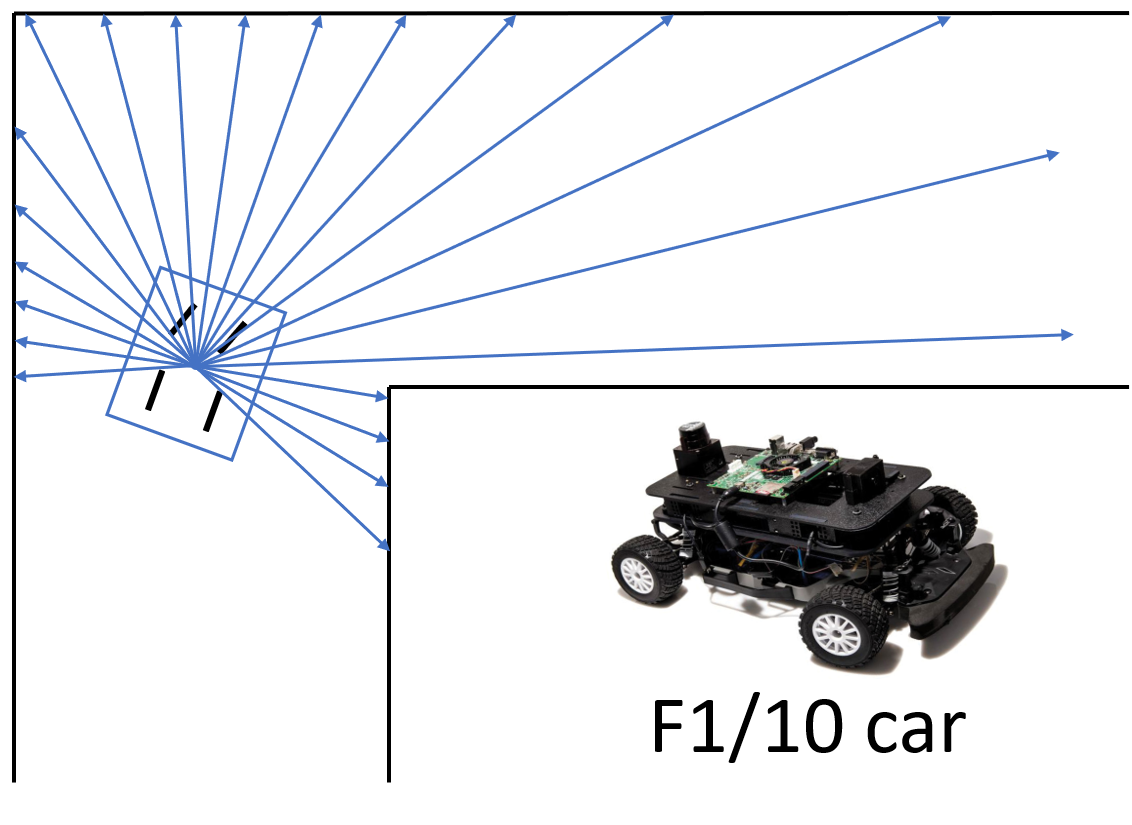}
}
\subfigure[Canonical scan at $x_k$][b]{%
\label{fig:f11_canonical}
\includegraphics[width=.24\linewidth]{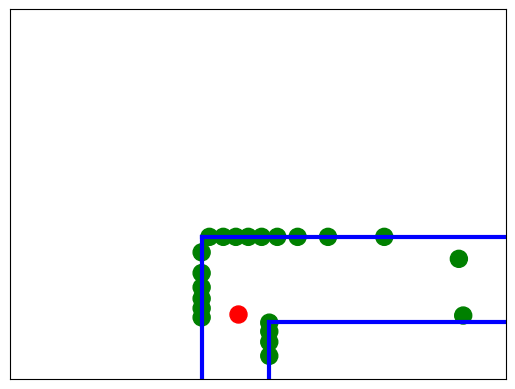}
}
\subfigure[Real scan at $x_k$][b]{%
\label{fig:f11_adversarial1}
\includegraphics[width=.24\linewidth]{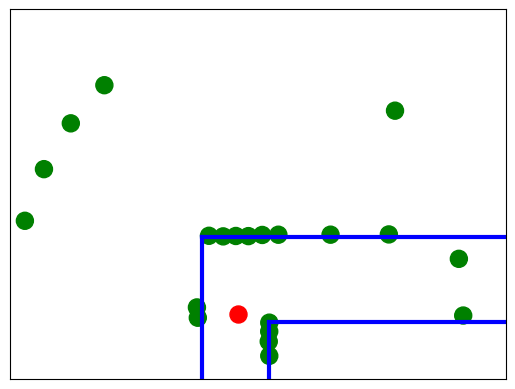}
}
\subfigure[Real scan at $x_{k+1}$][b]{%
\label{fig:f11_adversarial2}
\includegraphics[width=.24\linewidth]{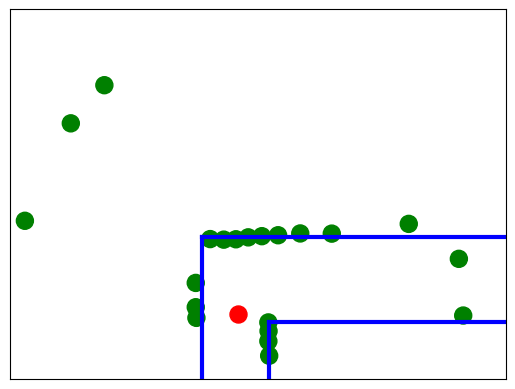}
}
\vspace{-10mm}
}\vspace{-4mm}  
\end{figure}

\vspace{-5px}
\subsection{The F1/10 Car: A LiDAR-Based System Operating Under Adversarial Noise}
The~\cite{f1tenth} is a popular autonomous racing platform. It exhibits a number of the challenges introduced by full scale autonomous cars, such as noisy measurements, adversarial agents, and fast-paced environments. We focus on the case of a single car navigating a square track using LiDAR measurements, as shown in Figure~\ref{fig:f110_overview}. In prior work, \cite{ivanov20a} used RL to train a number of controllers in a simulated environment and subsequently verified that these controllers are safe in the simulated environment. However, \cite{ivanov20a} also demonstrated that the controllers resulted in a number of accidents in the real world, as caused by noisy real data.

A canonical 21-dimensional LiDAR scan\footnote{A typical scan consists of 1081 rays, but we use a subsampled version to alleviate verification scalability challenges.} for a right-hand turn is illustrated in Figure~\ref{fig:f11_canonical}. Figures~\ref{fig:f11_adversarial1} and~\ref{fig:f11_adversarial2} show noisy real scans collected by \cite{ivanov20a} in a real environment similar to the simulated one; the car pose is the same (up to estimation error) in both environments. The real data contains reflected LiDAR rays (e.g., due to metallic surfaces), which manifest as perceived gaps in the environment. Such patterns are discontinuous and are classified as adversarial. Indeed, \cite{ivanov20a} showed that the number of accidents is significantly reduced if reflected surfaces are covered, thereby confirming the adversarial nature of LiDAR noise. Section~\ref{sec:modeling} presents a classifier-based approach for modeling adversarial noise for verification purposes.

\vspace{-10px}
\section{Modeling Framework}
\label{sec:modeling}

This section presents our modeling framework, including the canonical environment model and the two types of noise models. We also present a method for evaluating the quality of these components.


\vspace{-10px}
\subsection{Canonical Model}
 %
%
We focus on static environments where a first-principles model can describe the main features but may fail to capture environmental noise such as reflective surfaces and changing lighting conditions. Building a (canonical) first-principles model is particularly useful in such a setting as it enables the development of controllers in simulation. Once such a controller is built, deploying the system in the real environment requires only a ``de-noiser" for that specific environment, i.e., an additional component that is trained to translate real measurements to canonical ones. Thus, canonical models are by definition modular -- the same canonical model can be be composed with different noiser/de-noiser settings and can thus be used for different environments that are the same up to measurement noise (e.g.,~roads with various degrees of marking degradation). In future work, we will also consider the case where the canonical model is augmented with other agents observed in the real environment.

\vspace{-10px}
\subsection{Noise Model}
When developing a noise model for the purposes of verification, there are two main scalability-related requirements: 1)~models should be small and modular; 2)~models should not introduce significant spurious (unrealistic) noise, as that greatly complicates the verification task. Training a single model that satisfies both of these requirements is challenging, especially when one considers the types of noise illustrated in Section~\ref{sec:examples}. In particular, such a model would be exceedingly large for verification purposes and would likely introduce a large number of spurious behaviors in order to capture discontinuous noise (NNs are by definition continuous, so the only way to capture discontinuous dynamics is through including all in-between values as well). Thus, we argue that it is necessary to train a separate noise model for each type of noise. In what follows, we define each type of noise formally and describe a training procedure for the corresponding noise model.
\paragraph{Benign Noise.}
Intuitively, benign noise is noise that is easy to learn, both to generate and to be robust to. In their seminal paper, \cite{szegedy13} show that NN classifiers are robust to Gaussian noise, even if the noise results in more distorted images than adversarial noise. We observe a similar pattern in the context of the examples considered in this paper: continuous/benign noise is easier to learn than discontinuous/adversarial noise. Formally, we define benign noise as follows.
\begin{definition}[Benign Noise]
\label{def:benign}
Let $\Delta_x = g_n(x, g_c(x), \delta) - g_c(x)$ be the measurement noise introduced by the environment. We classify $\Delta_x$ as benign if it changes slowly across the state space $\mathcal{X}$:
\vspace{-7px}
\begin{equation*}
\forall \varepsilon_x>0\ \exists \varepsilon_{\Delta}>0 \text{ s.t. } \forall x,x' \in \mathcal{X},\ \|x-x'\|_\infty \le \varepsilon_x \Rightarrow \|\Delta_x - \Delta_{x'}\|_\infty \le \varepsilon_{\Delta}.
\end{equation*}
\end{definition}
%
Definition~\ref{def:benign} captures not only noise that is small in magnitude, but also noise that may be large in magnitude but is not changing across the state-space (e.g., a change in lighting conditions may result in large noise overall but is easy to learn as it is reasonably constant). This definition captures both types of noise considered in the MC case study, which introduce significant changes to the canonical image but are consistent across states. To train a generative model for benign noise given a training set ${\mathcal{D} = \{(x_i, y_i)\}}$, one can use a number of losses, e.g., mean squared error, binary cross entropy (BCE), or reconstruction loss as introduced by \cite{kingma13}. In the experiments, we use BCE with fully-connected NN models as this is the NN architecture supported by Verisig.

\paragraph{Adversarial Noise.}
Adversarial noise is high-frequency noise that may change quickly across the state space, as demonstrated in Figures~\ref{fig:f11_adversarial1} and~\ref{fig:f11_adversarial2}. For example, LiDAR rays are always reflected be some surfaces (or may be reflected by other surfaces only under certain angles). Such noise is similar to the adversarial noise for classification tasks discovered by \cite{szegedy13}.
%
\begin{definition} [Adversarial Noise.]
We classify $\Delta_x$ (introduced in Definition~\ref{def:benign}) as adversarial noise if there exists a subset $\mathcal{X}_a \subseteq \mathcal{X}$ such that $\forall x \in \mathcal{X}, \forall x_a \in \mathcal{X}_a,\ \|\Delta_x-\Delta_{x_a}\|_{\infty} > \varepsilon_a$ for some $\varepsilon_a>0$.
\end{definition}
%
As argued above, training a generative model for adversarial noise is challenging since we are effectively trying to capture a discontinuous space using a continuous model. Thus, we propose to use a classifier, $c_n: \mathcal{X} \times \mathcal{Y} \to \{0,1\}^m$, that determines whether a measurement (dimension) is corrupted by adversarial noise. If a certain dimension is flagged, then its value is reset to a default value, $y_a$, corresponding to the effect of that type of noise (e.g., reflected LiDAR rays are set to the maximum LiDAR range whereas pixels affected by glare can be reset to white). Formally,
\vspace{-3px}
\begin{align}
    g_n(x, g_c(x))^d = \left\{ \begin{array}{lll} g_c(x)^d &&\text{if } c_n(x, g_c(x))^d = 0 \\
      y_a &&\text{if } c_n(x, g_c(x))^d = 1,\end{array} \right.
\end{align}
\vspace{-7px}
$\\$
where $g_c(x)^d$ is the value in position $d$ of $g_c(x)$ (same for $g_n(x, g_c(x))^d$). Thus, $c_n$ learns which states are likely to result in adversarial noise. To train $c_n$, we transform the training set ${\mathcal{D} = \{(x_i, y_i)\}}$ into a classification set ${\mathcal{D}_c = \{((x_i, y_i), n_i)\}}$ where $n_i$ is a binary vector of labels indicating which dimensions of $y_i$ contains adversarial noise. To generate labels, we set $n_i^d = 0$ if $|y_i^d - g_c(x)^d| \le t_\Delta$ and $n_i^d = 1$, otherwise (for a user-defined threshold $t_\Delta$). If $\mathcal{D}$ is dense enough to identify the adversarial subset $\mathcal{X}_a$, then one could use a clustering algorithm to obtain more refined labels $n_i$. Since such a procedure was not necessary for our case studies, we leave this analysis for future work.
\subsection{Model Evaluation}
\label{sec:model_evaluation}
Since both noise modeling approaches discussed in this section are data-driven, a statistical evaluation is required as well. We evaluate the performance of the trained model on a labeled test set $\mathcal{D}_t = \{(x_i,y_i)\}$. Although a straightforward evaluation would be to compute the average error norm $\|y_i - g_n (x_i, g_c(x_i), \delta_i)\|$, such an evaluation may be misleading since element-wise distances in high-dimensional measurement spaces (e.g., images) are not semantically meaningful. For example, the model may capture the adversarial noise in an adjacent pixel/ray -- while such a model is not perfect, it may elicit the same overall system behavior. Thus, we evaluate the noise model on the downstream task (e.g., control or estimation), where distances are semantically meaningful.

We propose three evaluation metrics depending on the downstream task. The \emph{state-estimation-based metric} applies to systems with a state estimator (e.g., MC); in this case, we calculate the average error $\|h_e(y_i)-h_e(g_n(x_i, g_c(x_i),\delta_i))\|$ over the test set, where $h_e$ is a state estimator. Similarly, a \emph{control-based metric} calculates the average error of a controller $h_c$: $\|h_c(y_i) - h_c(g_n(x_i, g_c(x_i),\delta_i))\|$. Finally, a \emph{trajectory-based metric} compares the number of unsafe events recorded in the test set vs. the number of unsafe events recorded by simulating the system using the noiser model. The first two metrics can also be used to generate a PAC bound on the state estimation/control error. In future work, we will incorporate this bound in the verification problem as well.
\vspace{-10px}
\section{Verification}
\label{sec:verification}
\vspace{-3px}
Given models for all system components, we can perform verification using any verification tool for autonomous systems with NN components. We use our tool Verisig as it has shown great scalability on the examples considered in this paper. Verisig focuses on fully-connected NNs with smooth activations (e.g., sigmoid and tanh) and works by transforming the NN into a hybrid system that is then composed with the plant's dynamical system. The resulting hybrid system reachability problem is solved by the tool Flow* (\cite{chen12}), which uses Taylor Model approximations.
\begin{figure}[t]
  \centering
  \floatconts{fig:mc_eval}
{\caption{Evaluation of generative noise models on MC. Figures (a) and (b) show estimated position trajectories (using modeled vs. real observations) in two challenging noise scenarios. Figure (c) shows absolute position estimation errors over 1000 simulated trajectories.}}
{%
\centering
\subfigure[Low contrast ($\delta_c=0.3$)][b]{%
\label{fig:mc_eval_contrast}
\includegraphics[width=.31\linewidth]{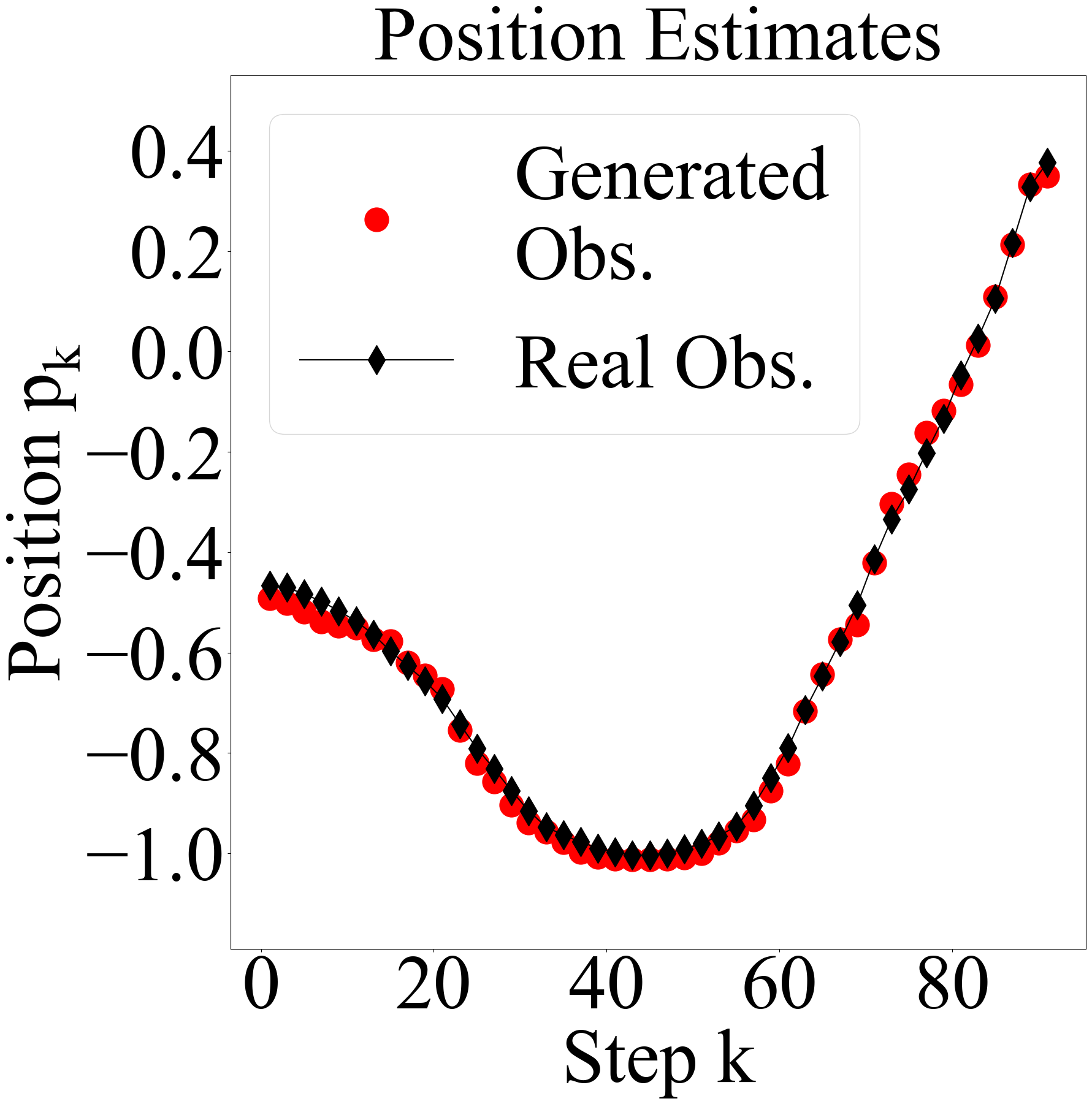}
}
\subfigure[High blur ($\delta_b=0.3$)][b]{%
\label{fig:mc_eval_blur}
\includegraphics[width=.33\linewidth]{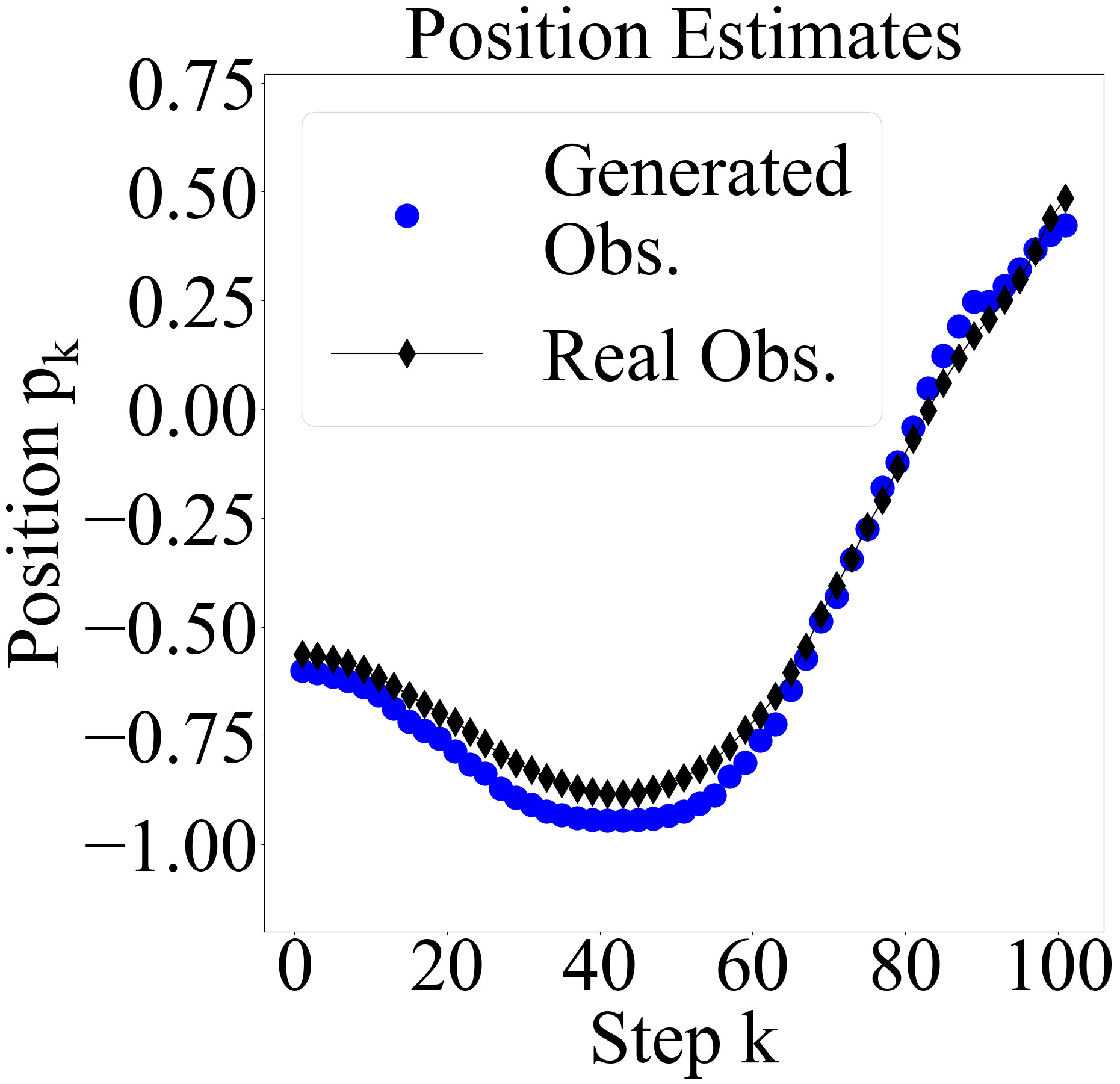}
}
\subfigure[Histogram of estimate errors][b]{%
\label{fig:mc_eval_hist}
\includegraphics[width=.31\linewidth]{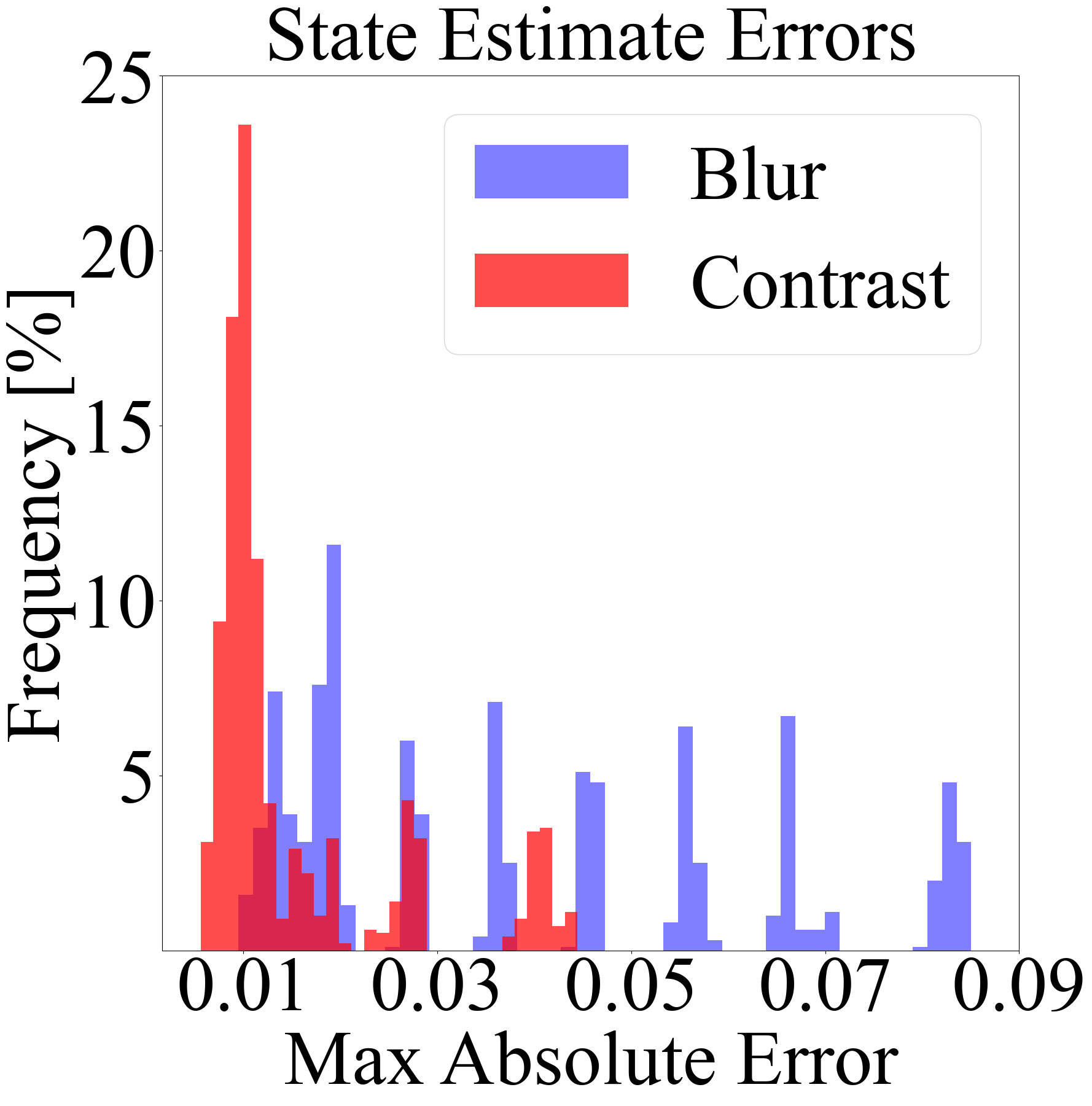}
}
\vspace{-7mm}
}\vspace{-4mm}  
\end{figure}

\vspace{-10px}
\section{Case Studies}
\vspace{-3px}
This section presents two case studies, based on the motivating examples from Section~\ref{sec:examples}. 
A more detailed description is available in the appendices.

\label{sec:case_studies}
\subsection{Mountain Car}
\vspace{-5px}
As described in Section~\ref{sec:examples}, in MC the controller observes an image and ground-truth velocity. The car's dynamics are: ${p_{k+1} = p_k + v_{k}}$, ${v_{k+1} = v_k + 0.0015u_k - 0.0025*cos(3p_k)}$, where $p$ and $v$ are position and velocity, respectively, and $u \in [-1,1]$ is thrust. The initial condition $p_0 \in [-0.6,-0.4]$ is at the bottom of the mountain. We would like to verify the system's robustness to two types of noise, contrast and blur. Since images are produced by a black-box simulator, we train a generative canonical observation model, described next. The control pipeline consists of two parts: 1) a state estimator, $h_s$, that estimates the car position from an image (explained next); 2) a controller, $h_c$, that takes the position estimate and ground-truth velocity as inputs. The controller is a pre-trained and pre-verified (for $p_0 \in [-0.59, 0.4]$) NN, borrowed from \cite{ivanov19}.

\vspace{-5px}
\paragraph{Model Training.}
To train all models, we create a training set $\mathcal{D} = \{(p_i,y_i^c, y_i^{ct}, y_i^b)\}$ of 82000 examples. We evenly sample 2000 positions and canonical images (grayscaled and downsampled to $20 \times 30$), $y_i^c$; for each one, we create 40 images with varying contrast, $y_i^{ct}$, and blur, $y_i^b$. 
The state estimator, $h_e: \{y_i^c, y_i^{ct}, y_i^b\} \mapsto p_i$, is trained using least squares on all images. The canonical model, $g_c: p_i \mapsto y_i^c $, is trained using BCE.
The contrast model, ${g_{ct}: y_i^c \times \delta_{c,i} \mapsto y_i^{ct}}$, and the blur model, $g_{b}: y_i^c \times \delta_{b,i} \mapsto y_i^{b}$, map a canonical image to a noised image, for a range of contrast, $\delta_{c,i}\in [0.3, 2.0]$, and blur, $\delta_{b,i} \in [0, 0.3]$.
%
All models are fully-connected NNs with tanh activations with the following shapes (notation $[a_1, \dots, a_n]$ means the NN has $n$ layers with $a_i$ neurons each; if $a_i$ are the same, we write $n\times a$): 1) $h_e: 3\times 20$; 2) $g_c: [50,100]$; 3) $g_{ct}: 1\times 50$; 4) $g_b: 1\times 50$.


\vspace{-5px}
\paragraph{Model Quality.}
To evaluate models, we use the state-estimation-based metric discussed in Section~\ref{sec:model_evaluation}. We compare state estimates on 2000 test trajectories (with varying noise levels) using: 1)~ground truth images vs. 2)~images produced by the data-driven model.
Figures~\ref{fig:mc_eval_contrast} and~\ref{fig:mc_eval_blur} show estimated trajectories under the most challenging scenarios of low contrast and high blur conditions, respectively. As can be seen in the figures, both noise models result in very low difference in estimation performance. Figure~\ref{fig:mc_eval_hist} summarizes the absolute errors over all trajectories. All errors are bounded within a small range; although blur errors have a longer tail, the error is bounded by 0.09, which corresponds to 5\% of the total position space. 
It is important to note that one can even compose the blur and contrast models to achieve more complex noise patterns. Although not shown in the interest of space, the compositional model achieves similar performance
on restricted domain for noise parameters (namely, $\delta_c \in[.5, 2]$ and $\delta_b \in [0, 0.15]$). 
Based on these promising results, we conclude that the noise models are sufficiently accurate so that a verification result is meaningful.
\begin{table}[t]
    \centering
    \scriptsize
    \begin{tabular}{|l|c|c|c|c|c|}
    \hline
        \multicolumn{1}{|c|}{System} & \multicolumn{1}{c|}{ Initial Conditions} &  \multicolumn{1}{>{\centering\arraybackslash}p{10mm}|}{Total Intervals}&  \multicolumn{1}{>{\centering\arraybackslash}p{10mm}|}{Avg. Time [s]}  &    \multicolumn{1}{>{\centering\arraybackslash}p{15mm}|}{Avg. NN Time [s]} &  \multicolumn{1}{>{\centering\arraybackslash}p{10mm}|}{Avg. Branches}  \\
        \hline
        MC: Contrast & $p_0 \in [-.59, -.4]$ $\delta_{c} \in [0.3, 2]$ & 816 & 48396 & 26259 & 1.26  \\
        \hline
        MC: Blur & $p_0 \in [-.59, -.4]$ $\delta_b \in [0, 0.3]$ & 2057 & 62476  & 29702 & 1.02 \\
        \hline
        MC: Composite & $p_0 \in [-.55, -.4]$ $\delta_c \in [0.5, 2] $
        $\delta_b \in [0, 0.15] $ & 1272 & 61395 & 32894 & 1.38 \\
        \hline
        F1/10 & $x_{1,0} \in [-0.1, 0.1]$ $x_{2,0} \in [1, 1] $
        $\theta_{0} \in [0, 0] $ & 8236 & 10733 & 4482 & 6.88 \\
        \hline
    \end{tabular}
    \vspace{-7px}
    \caption{Verification results for both case studies. The last column shows the number of branches in the composed system, e.g., due to the adversarial noise classifier possibly outputting two values due to uncertainty. Branches present a scalability challenge as each one needs to be verified separately.}\label{tab:verification}
    \vspace{-6.5mm}
    
\end{table}

\vspace{-5px}
\paragraph{Verification.}
We verify that the car goes up the hill (with a reward of at least 90, which was the property verified by \cite{ivanov19} for the original controller) for a range of initial positions and noises, as shown in Table~\ref{tab:verification}. All ranges were verified safe. The verification was carried out in parallel using sub intervals (of size $0.00025$) of the initial state space and the noise range. If verification of any given interval did not terminate within 24 hours (we used a server with 95 cores and 500GB of RAM) the process was restarted with five sub-intervals (and restarted again if necessary).

\begin{wraptable}{r}{0.55\textwidth}
\vspace{-20px}
\scriptsize
\centering
\begin{tabular}{|r|c|c|c|c|c|}
    \hline
        \multicolumn{6}{|c|}{\textbf{Noiser Training Results}}\\
            \hline
        Model Size & 3x100 & 4x100 & 5x100  &  6x100 & 6x200\\
        \hline
        WSA [\%] & 61& 77 & 87 & 98 & 100 \\
        \hline
        Control Err. & 3700 & 2312 & 1485 & 774 &  594 \\
        \hline
        \multicolumn{6}{|c|}{\textbf{Safe Trajectory Distribution [\%]}} \\
        \hline
        \textbf{Real Data} & \multicolumn{5}{|c|}{\textbf{Simulated Data Using Noiser Model}} \\
        \hline
        Controller $c_1$:  \textbf{80} & 88 & 83 & 88 & 82 &  97 \\
        \hline
        Controller $c_2$: \textbf{90}  & 75 & 80 & 92 & 77 &  74 \\
        \hline
        Controller $c_3$: \textbf{90}  & 85 & 78 & 98 & 91 &  91 \\
        \hline
        Controller $c_4$: \textbf{0}  & 29 & 29 & 29 & 32 &  27 \\
        \hline
    \end{tabular}
    \vspace{-5px}
    \caption{Training results and crash distributions for different noise models. Pre-trained controllers and real data were borrowed from \cite{ivanov20a} (controllers $c_1$, $c_2$, $c_3$, and $c_4$ correspond to controllers ``DDPG,64,C3", ``TD3,64,C2", ``TD3,128,C2", and ``DDPG,128,C3").}
    \label{tab:f1_train_crash}
    \vspace{-10px}
\end{wraptable}
\subsection{The F1/10 Car}
\vspace{-5px}
The F1/10 case study is significantly more challenging than MC since it is based on real data, and it includes substantial adversarial noise. We use real data traces collected by \cite{ivanov20a} that were designed to evaluate the simulation-to-reality (sim2real) gap between the modeled and the real system. We aim to bridge the sim2real gap by training a data-driven noise model, as well as a ``denoiser" that filters adversarial noise in order to use the original controller. We first train high-quality noise and denoiser models, before verifying the car safely navigates the track.


\vspace{-5px}
\paragraph{Model Training.}
Since the data traces do not contain ground truth car poses, we develop a state estimator using a particle filter (\cite{thrun05}), based on the bicycle and LiDAR models in \cite{ivanov20a}. Using the state estimates, we create a training set ${\mathcal{D} = \{(\mathbf{x}_i, y_i)\}}$, where $\mathbf{x}_i = (x_{1,i}, x_{2,i}, \theta_i)$ denotes the car's $(x,y)$-coordinates and orientation, respectively, and $y_i$ is the corresponding LiDAR scan. 
To train the LiDAR noise model, we label each LiDAR scan according to the procedure in Section \ref{sec:modeling}. 
Given a canonical scan $g_c(\mathbf{x}_i)$, we set ${n_i^d = 1}$ if ${|g_c(\mathbf{x}_i)^d - y_i^d| > t_{\Delta}^*}$, and ${n_i^d = 0}$, otherwise.\footnote{The threshold is chosen to minimize the error between noised canonical scans and real scans over the training data.}
Using the labeled training set ${D_l = \{((g_c(\mathbf{x}_i),\mathbf{x}_i), n_i)\}}$,
we train classifiers of increasing sizes (listed in Table~\ref{tab:f1_train_crash}) using weighted BCE (to account for class imbalance between adversarial and non-adversarial examples).
Training multiple noise models allows us to: 1) demonstrate that the task is feasible; 2) identify the largest model that can be handled by Verisig.
Finally, we train a $1\times 100$
fully-connected denoiser NN, $h_d: y_i \mapsto \hat{y}_i$, on pairs $(g_c(\mathbf{x}_i), y_i)$.

\begin{figure}[t]
  \centering
  \floatconts{fig:f1_eval}
{\caption{Control-based evaluation of F1/10 noise model, using controller $c_2$. Figures ($a$) and ($b$) compare control outputs on two trajectories for canonical, noised, and real measurements. Figure ($c$) summarizes the maximum and mean errors over all 115 real trajectories. }}
{%
\centering
\subfigure[Moderate max error][b]{%
\label{fig:f1_model_eval_control}
\includegraphics[width=.295\linewidth]{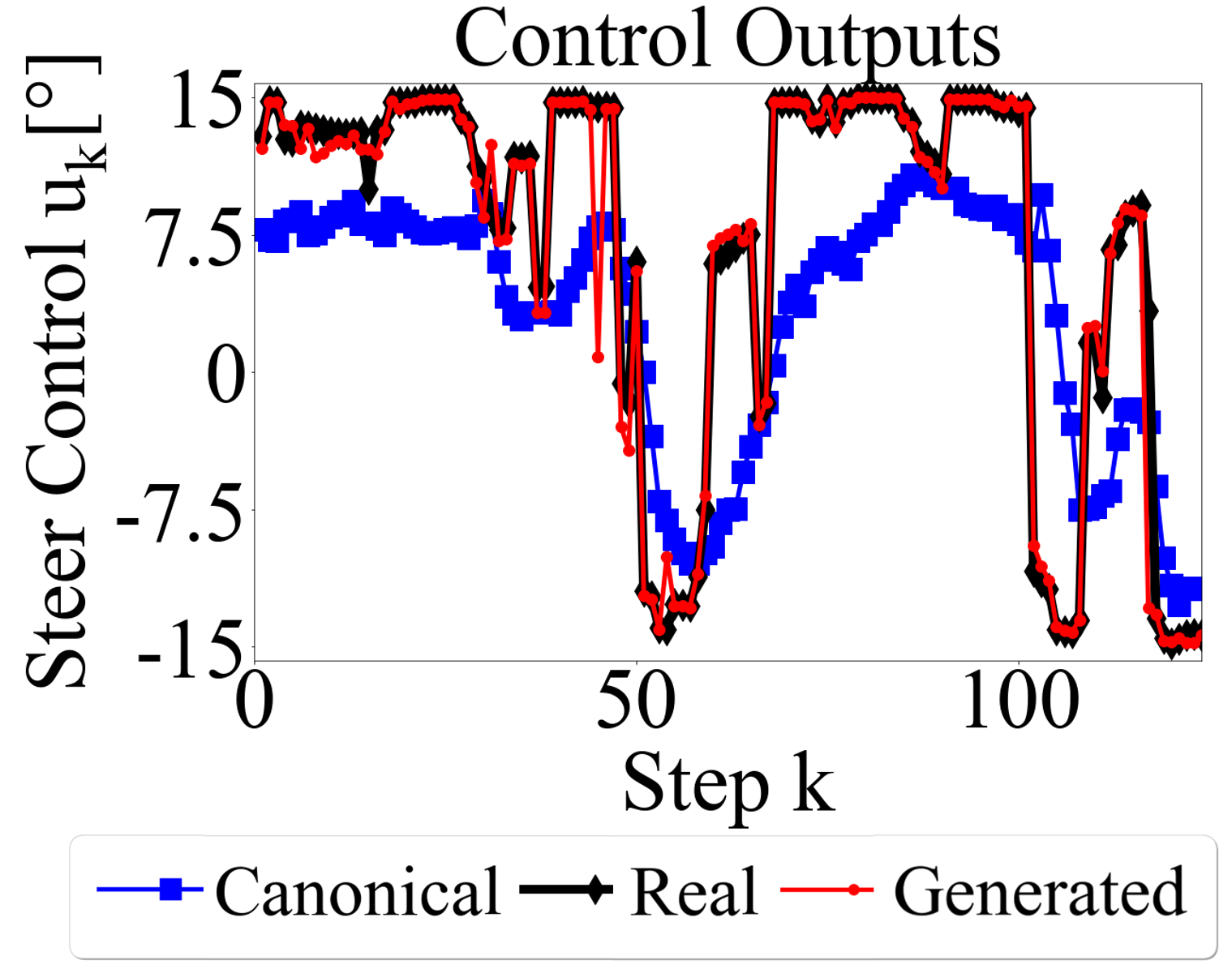}
\hspace{3mm}
}
\subfigure[High max error at step 47][b]{%
\label{fig:f1_model_eval_control2}
\includegraphics[width=.295\linewidth]{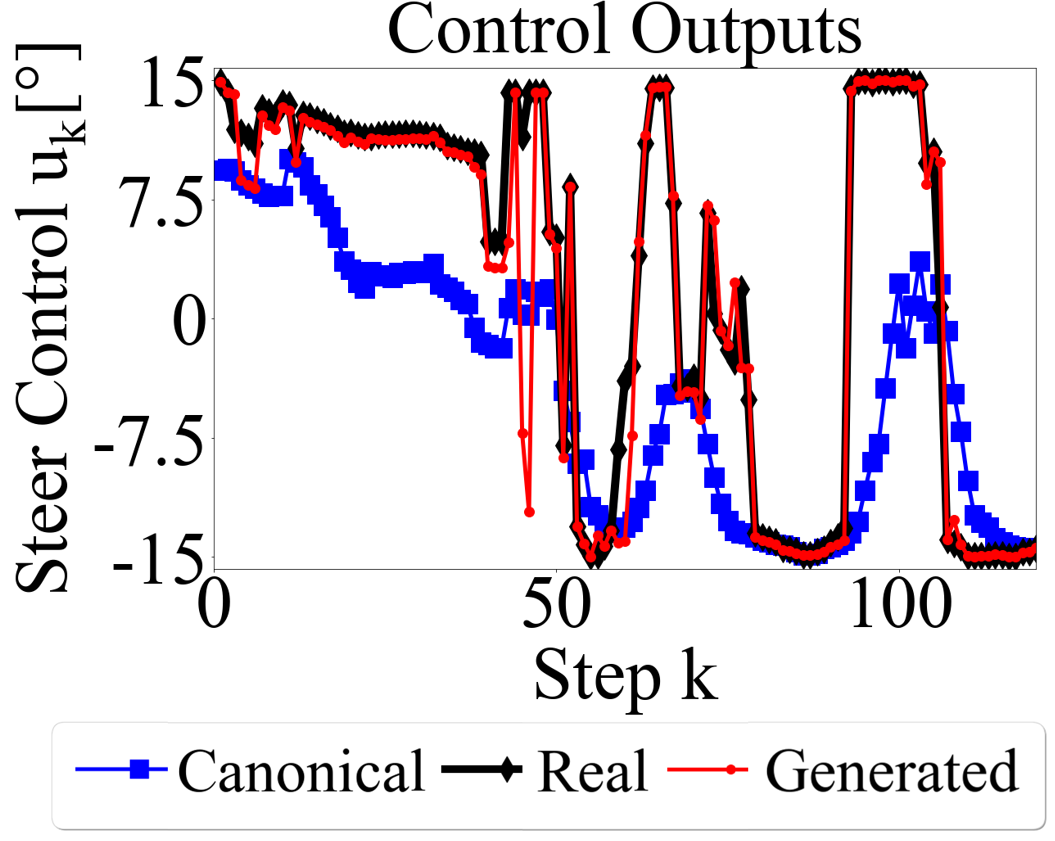}
}\hspace{3mm}
\subfigure[Histogram of control errors][b]{%
\label{fig:f1_eval_control_hist}
\includegraphics[width=.31\linewidth]{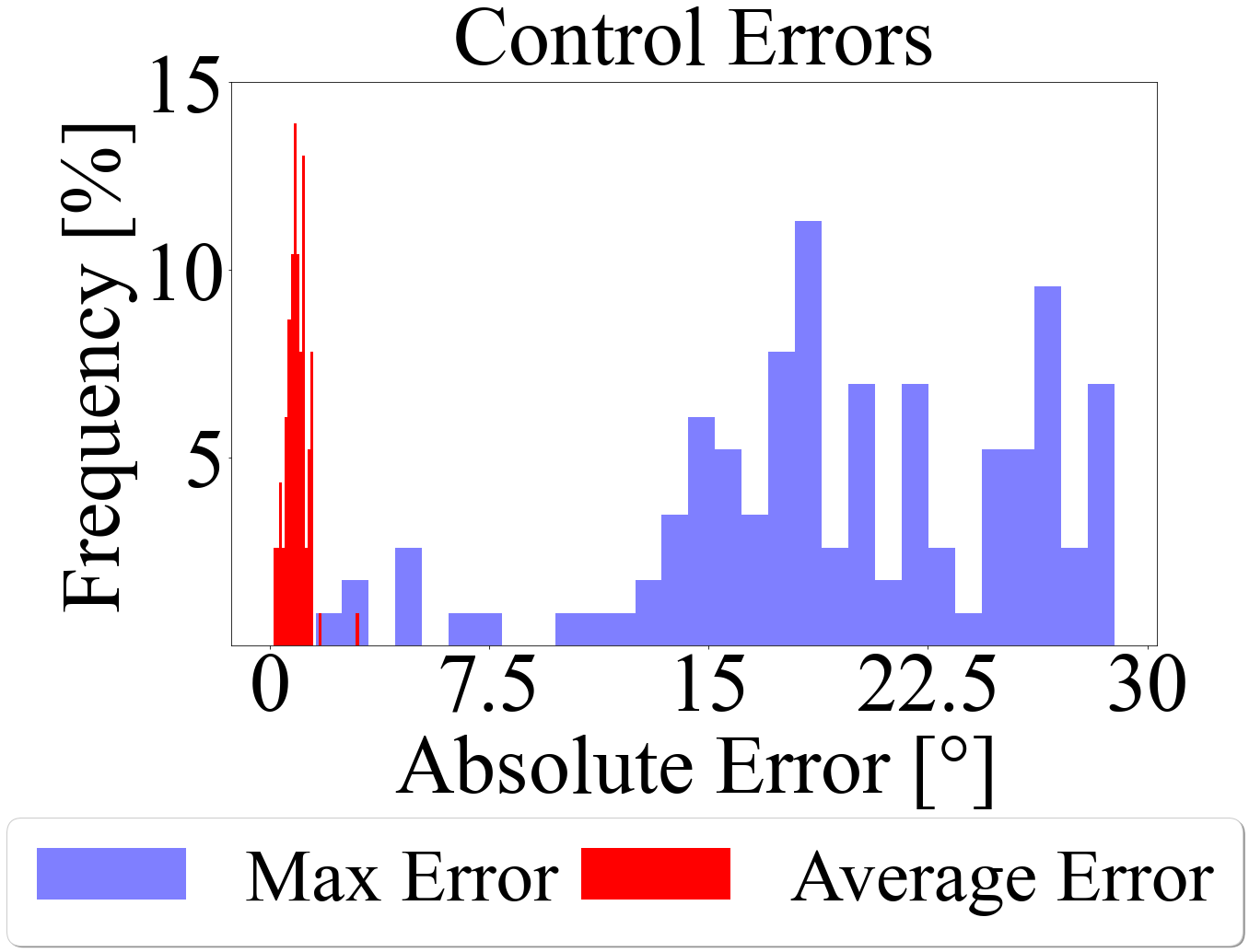}
}

\vspace{-6mm}
}\vspace{-5mm}  
\end{figure}

\begin{figure}[t]
  \centering
   
  \floatconts{fig:f1_trajectories}
{\caption{Simulated trajectories under the 5x100 noiser model vs. real trajectories for four different controllers (without using a denoiser). Bold trajectories indicate those that ended in a crash.}\label{fig:f1_simvreal}}
{%
\centering
\subfigure[Controller $c_1$ ][b]{%
\label{fig:f1_simvreal1}
\includegraphics[width=.19\linewidth]{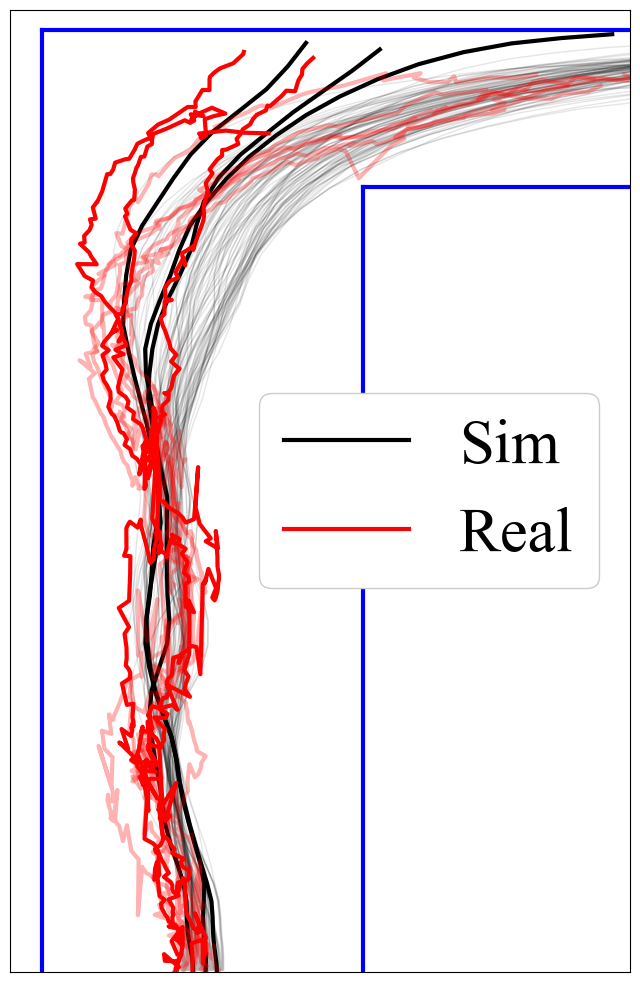}
}
\hfill
\subfigure[Controller $c_2$ ][b]{%
\label{fig:f1_simvreal3}
\includegraphics[width=.19\linewidth]{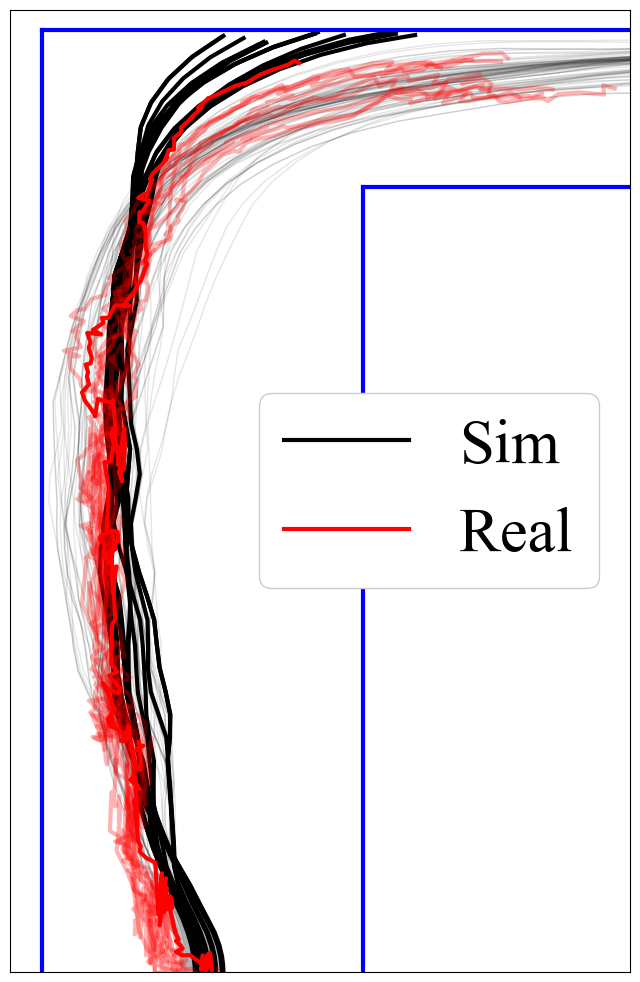}
}
\hfill
\subfigure[Controller $c_3$][b]{%
\label{fig:f1_simvreal4}
\includegraphics[width=.19\linewidth]{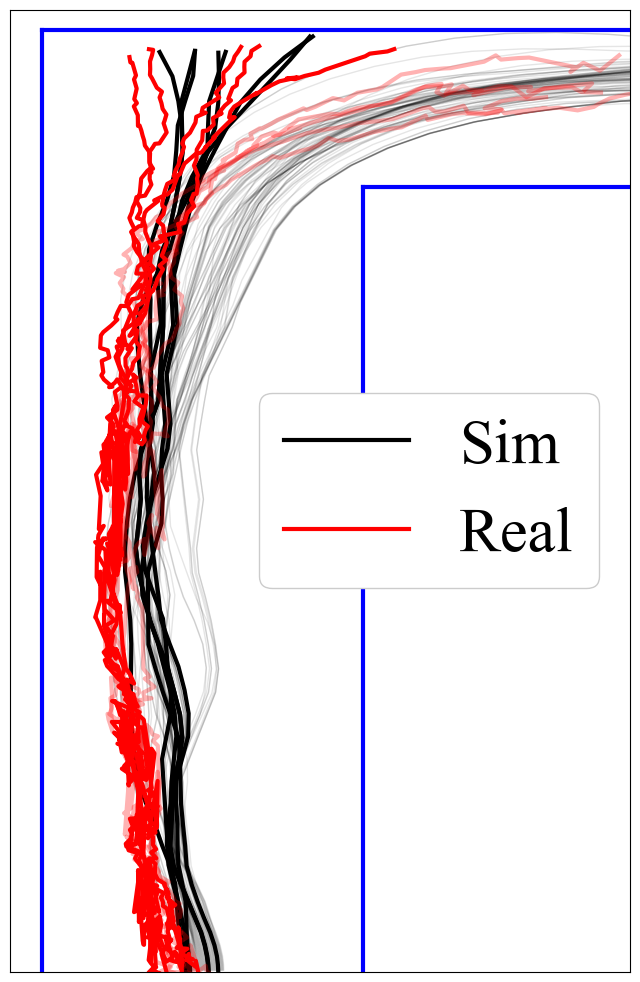}
}
\hfill
\subfigure[Controller $c_4$][b]{%
\label{fig:f1_simvreal2}
\includegraphics[width=.19\linewidth]{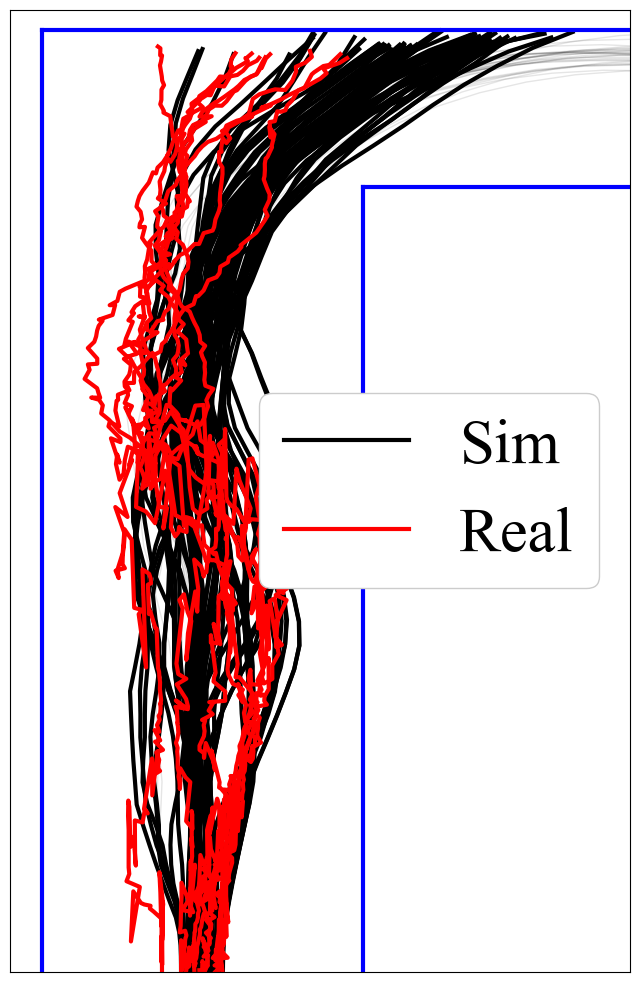}
}
\vspace{-7mm}
}\vspace{-5mm}  
\end{figure}

\vspace{-5px}
\paragraph{Model Quality.}
Since training data is limited (115 trajectories), we train on all data and do not have a test set (in future work, we will perform an exhaustive cross-validation study). To get around this issue, we use both the control-based metric and the trajectory-based metrics discussed in Section~\ref{sec:model_evaluation}. Table~\ref{tab:f1_train_crash} shows an evaluation of all trained noise models according to whole scan accuracy (WHA); WHA is preferred to individual accuracy as even a single adversarial ray can significantly affect control output. The table shows that WHA is directly proportional to lower error between control outputs on the model vs. real scans. To evaluate the model according to the control-based metric, we use four pre-trained and pre-verified controllers from \cite{ivanov20a}, labeled $c_1,\dots, c_4$. Figures~\ref{fig:f1_model_eval_control} and~\ref{fig:f1_model_eval_control2} show the control outputs, when using $c_2$, on two example real trajectories. We note that the adversarial noise model is very accurate, especially compared to the canonical model, which produces significantly different control outputs. Furthermore, Figure \ref{fig:f1_eval_control_hist} shows the maximum and average control errors over all trajectories. Unlike MC, the error is not always bounded and is sometimes equal to the full control range of 30 degrees.
At the same time, the average error is quite small, indicating great model performance on average.

Finally, we evaluate the model in terms of the trajectory-based metric, which effectively serves as a test set since the simulated trajectories deviate somewhat from the car states observed in real data. The bottom rows of Table~\ref{tab:f1_train_crash} compare the number of crashes observed in the real data vs. the number of crashes observed in simulation for the same controllers (without using the denoiser). Not only are the number of crashes very close, but we also demonstrate qualitatively in Figure~\ref{fig:f1_simvreal} that the crashing trajectories observed in the real experiments look very similar to the simulated ones when the data-driven noise model is used. These results suggest that the learned LiDAR noise model is a good approximation of real LiDAR noise and is thus an appropriate model for verification purposes.

\vspace{-7px}
\paragraph{Verification.} Having established the accuracy of the data-driven noise model, we now verify that the car can safely navigate the noisy environment when also equipped with a denoiser. Using controller $c_2$ and the 5x100 noiser model, we verify a that if the car starts in the range $x_{1} \in [-0.1, 0.1]$ in the middle of the hallway, then it can safely navigate the right-hand turn (a similar property was verified by \cite{ivanov20a}). The verification statistics are reported in Table~\ref{tab:verification}. Similar to MC, we split the initial range into sub-intervals and iteratively verify the safety of each initial interval.

\vspace{-16px}
\section{Conclusion}
\vspace{-7px}
This paper addressed the problem of data-driven modeling and verification of perception-based autonomous systems. We proposed a compositional modeling approach by using a first-principles canonical model, coupled with a data-driven model capturing the real environment's noise, both in the case of benign and adversarial noise. We presented two cases studies illustrating the challenges of training and verifying such a compositional model. Future work includes performing an experimental evaluation that demonstrates the benefit of verification on a real platform. Futhermore, we will investigate the development of noise models that include additional agents in the environment.


\bibliography{references.bib}
\newpage

\appendix

\section{Mountain Car}
This section provides a detailed explanation of the mountain car (MC) model and the verification task.

\subsection{Dynamics}
The MC dynamics are as follows:
\begin{align*}
p_{k+1} &= p_k + v_{k}\\
v_{k+1} &= v_k + 0.0015u_k - 0.0025*cos(3p_k),
\end{align*}
where $p_k$ and $v_k$ are position and velocity, respectively, with $p_0 \in [-0.6, -0.4]$. Note that $v_k$ is constrained to be within $[-0.07, 0.07]$ and $p_k$ is constrained to be within $[-1.2, 0.6]$; this means that the MC model is actually a hybrid system when those constraints are reached. The inputs $u_k \in [-1,1]$ are thrust.

\subsection{Controller}
The controller, $h_c$, is a $2\times 16$ NN with sigmoid activations in the hidden layers and tanh activation on the output neuron. This controller was verified to be safe by \cite{ivanov19}, i.e., it reaches the goal with a reward of at least 90. Note that this controller takes position and velocity as input.

\subsection{Training Set}
In this paper we focus on the case where the control pipeline does not observe position and velocity but rather observes an image and ground-truth velocity. Furthermore, we consider two types of noise, contrast and blur, as illustrated in Figure~\ref{fig:mc_examples}. In order to train all models described below, we build a training set  $\mathcal{D} = \{(p_i,y_i^c, y_i^{ct}, y_i^b)\}$ of 82000 examples. We evenly sample 2000 positions and corresponding canonical images (grayscaled, normalized to [0,1], and downsampled from the original $400 \times 600 \times 3$ to $20 \times 30$), $y_i^c$, as well as 20 contrasted, $y_i^{ct}$, and 20 blurred, $y_i^b$, images per position. Specifically, for each position $p_i$, the contrasted and blurred images are sampled evenly with $\delta_{c}\in [0.3, 2.0]$ and $\delta_{b} \in [0, 0.3]$. Contrast is added using the Python Image Library (PIL) ImageEnhance module where $\delta_{c}=0$ produces a solid gray image, $\delta_{c}=1$ produces the original image, and $\delta_{c}>1$ produces a higher contrast version of the original image \cite{clark2015pillow}. Blur is added as follows: $y_i^{b} = 0.5g_c(p_i-\delta_{b,i}) + y_i^c +0.5g_c(p_i+\delta_{b,i})$, i.e.,~a canonical image with a lighter overlay of left and right shifted images. Blurred images are then normalized to [0, 1].

\subsection{State estimator}
The state estimator $h_e: \{y_i^c, y_i^{ct}, y_i^b\} \mapsto p_i$ takes an image as input and outputs position. The state estimator is trained using least squares on all images in $\mathcal{D}$ and on a set of 100000 synthetic composite images $\{g_b(g_c(y_i^c, \delta_{c,i}),\delta_{b,i}) \}$. The state estimator is $3 \times 20$ NN with tanh activations on the hidden layers and a tanh activation on the output neuron. The output neuron is normalized to output only valid positions, values in the range $[-1.2, 0.6]$.

\subsection{Canonical Model}
Since the MC simulator is a black-box simulator, we do not have a first principles canonical model in this case. Thus, we use the canonical images in $\mathcal{D}$ to train a $[50,100]$ generative model, $g_c: p_i \mapsto y_i^c $, that approximates the simulator. Figure ~\ref{fig:mc_canonical_generative} demonstrates the performance of the generative canonical model according to the state-estimation-based metric. As can be seen in the figures, the generative canonical model is a very good approximation of the simulator. The generative canonical model is trained using BCE.

\begin{figure}[t]
  \centering
  \floatconts{fig:mc_canonical_generative}
{\caption{Evaluation of generative canonical model on MC. Figures (a) and (b) show estimated position trajectories (using modeled vs. real observations) in simulations. Figure (c) shows absolute position estimation errors over 100 simulated trajectories.}}
{%
\centering

\subfigure[Trajectory ($p_0=-0.53$)][b]{%
\label{fig:mc_eval_canonical1}
\captionsetup{justification=centering}
\includegraphics[width=.315\linewidth]{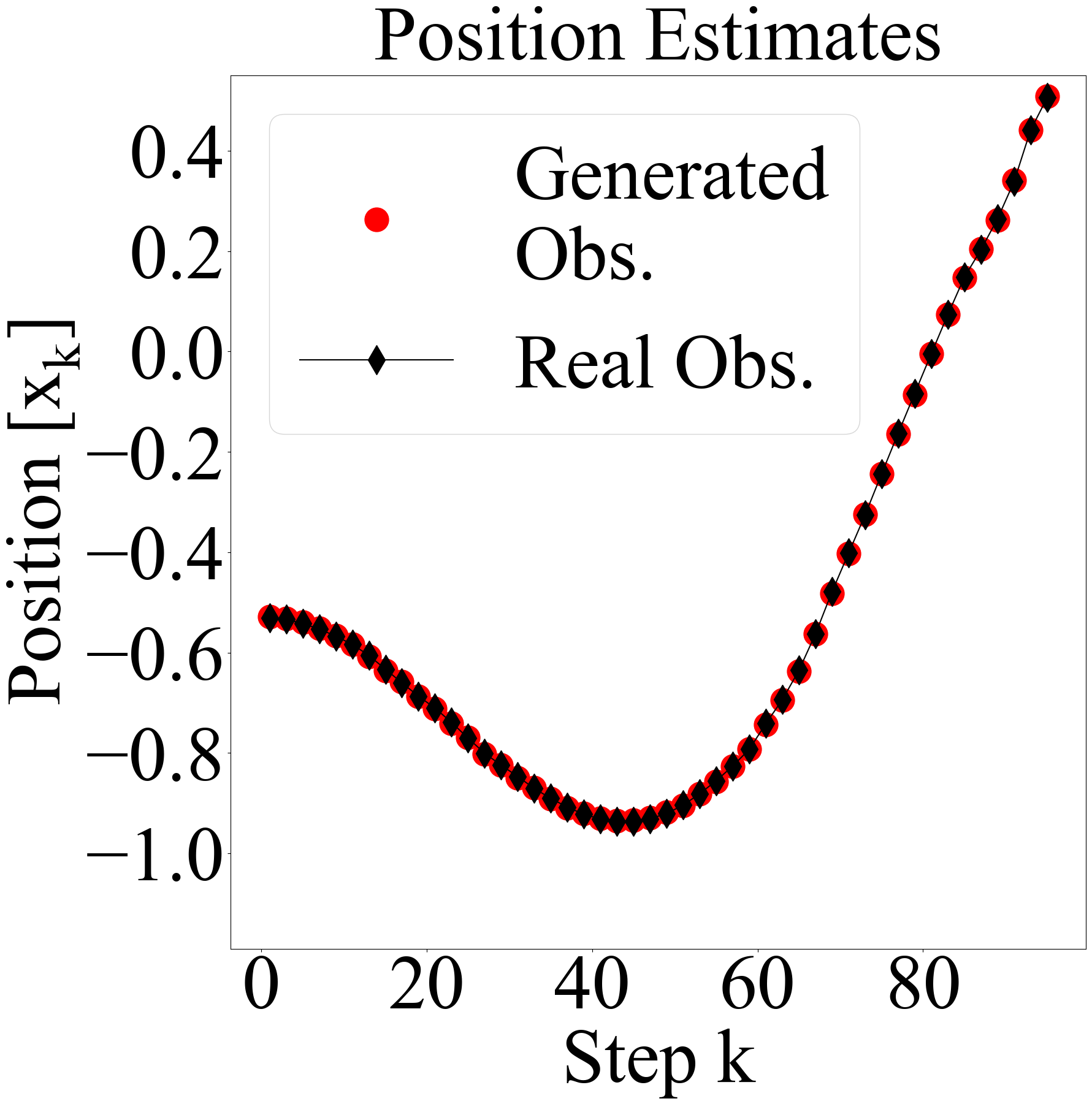}
}
\subfigure[Trajectory ($p_0=-0.4$)][b]{%
\label{fig:mc_eval_canonical2}
\includegraphics[width=.315\linewidth]{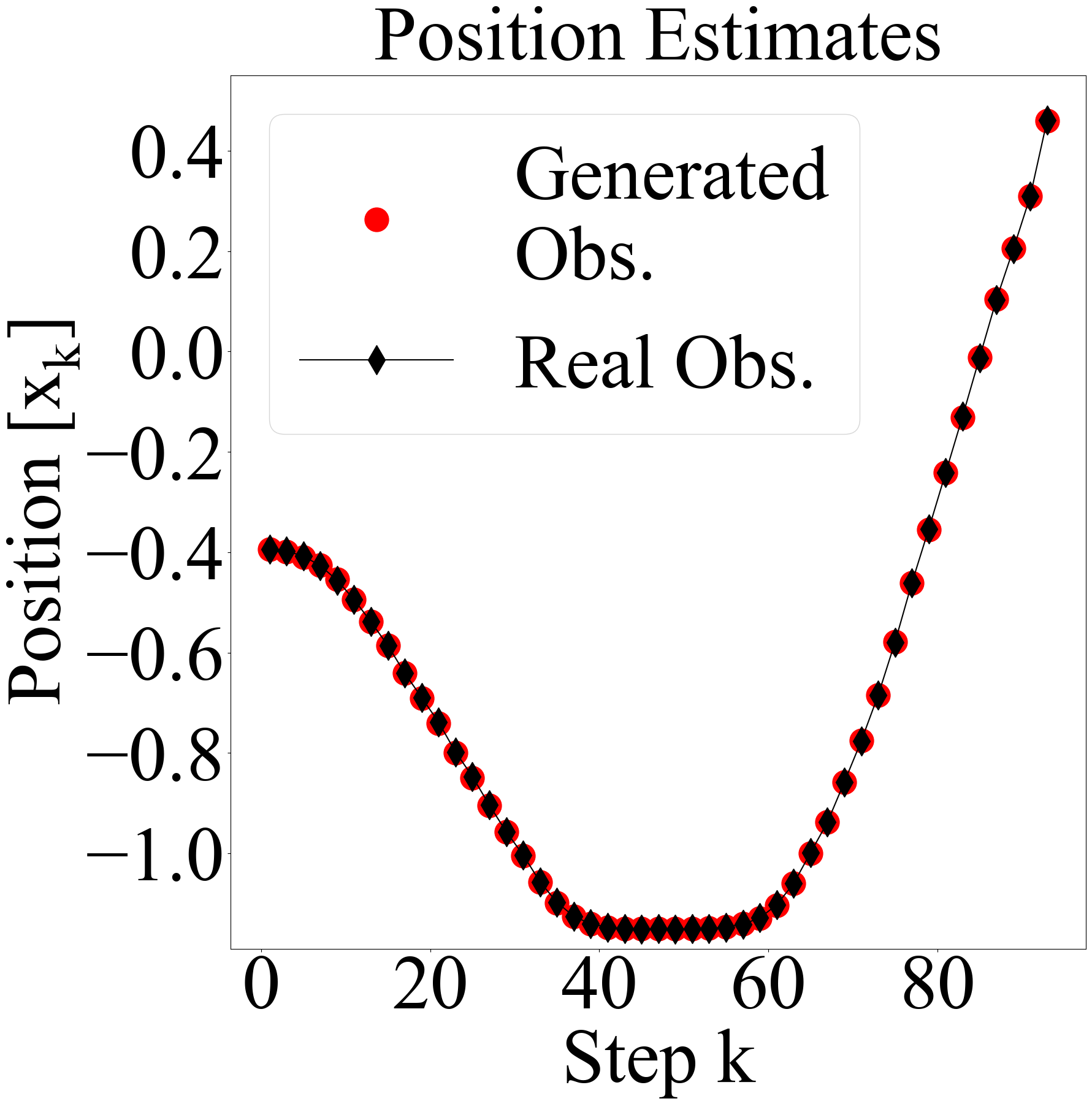}
}
\subfigure[Histogram of estimate errors][b]{%
\label{fig:mc_eval_canonical_hist}
\includegraphics[width=.32\linewidth]{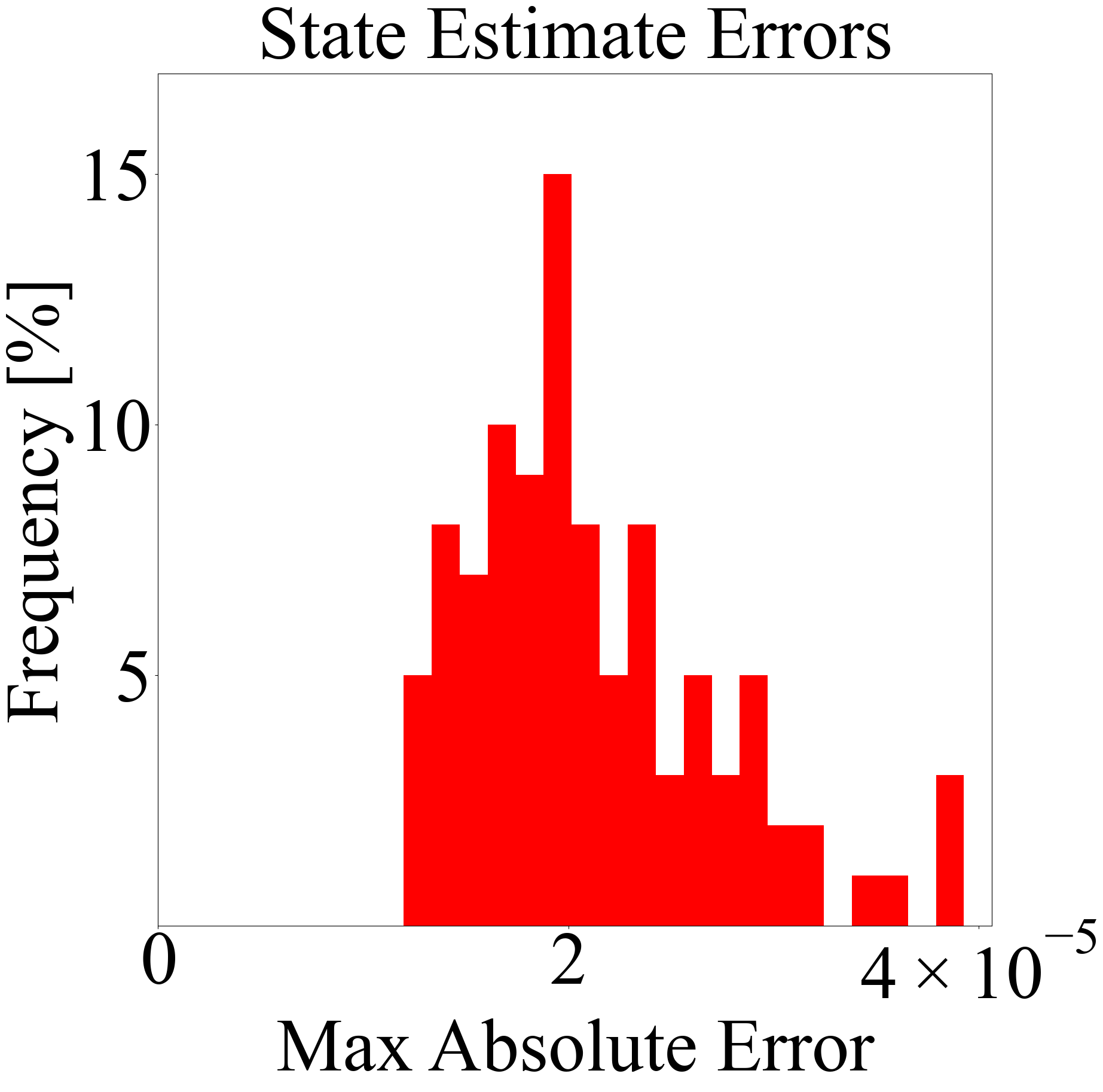}
}
\vspace{-7mm}
}\vspace{-4mm}  
\end{figure}

\subsection{Contrast Noise Model}
The contrast model, ${g_{ct}: y_i^c \times \delta_{c,i} \mapsto y_i^{ct}}$, maps a canonical image to a noised image for a range of contrast intensities, $\delta_{c,i}\in [0.3, 2.0]$. The contrast model is $1 \times 50$ and is trained using BCE and weight decay.

\begin{figure}[t]
  \centering
  \floatconts{fig:mc_eval_contrast2}
{\caption{Additional evaluation of generative contrast models on MC. Figures show estimated position trajectories (using modeled vs. real observations) in simulations. Figure ($a$) shows a trajectory under low contrast, ($b$) shows a trajectory under little contrast change from the canonical environment, and ($c$) shows a trajectory under high contrast. }}
{%
\centering

\subfigure[Low contrast ($\delta_c=0.3$)][b]{%
\label{fig:mc_eval_contrast2_1}
\captionsetup{justification=centering}
\includegraphics[width=.315\linewidth]{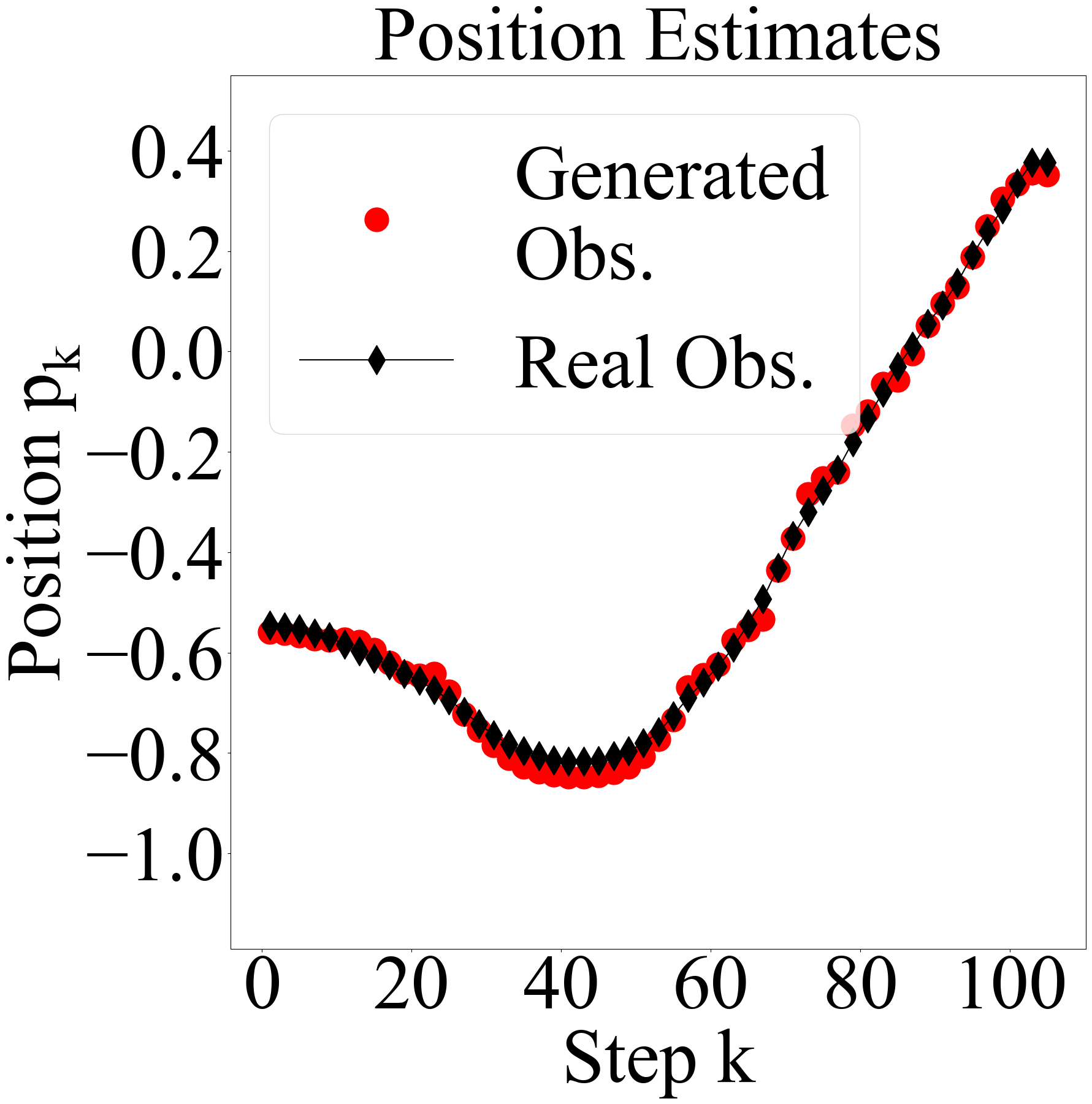}
}
\subfigure[Near canonical contrast ($\delta_c=1.06$)][b]{%
\label{fig:mc_eval_contrast2_2}
\includegraphics[width=.315\linewidth]{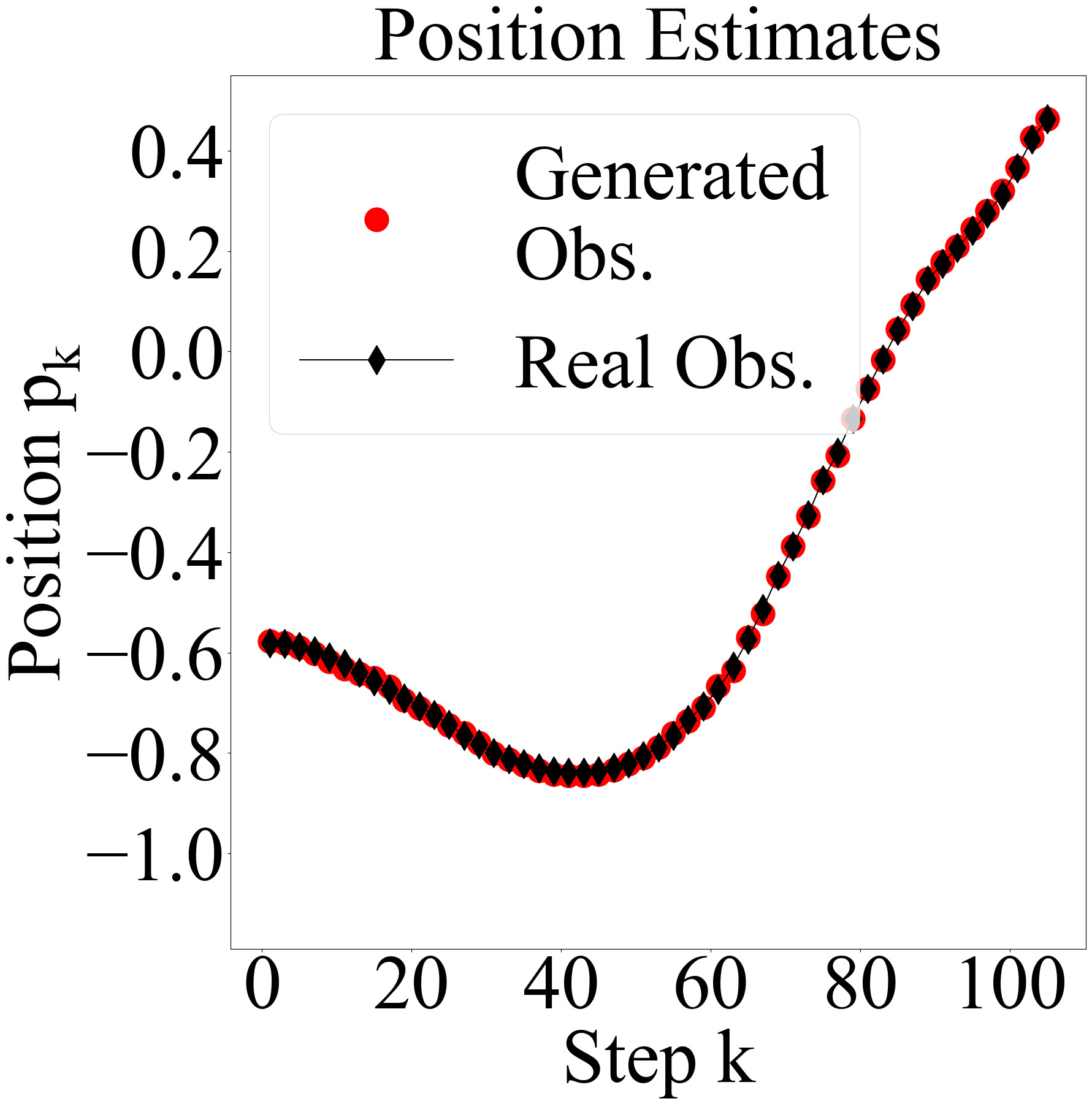}
}
\subfigure[High contrast ($\delta_c=2.0$)][b]{%
\label{fig:mc_eval_contrast2_3}
\includegraphics[width=.32\linewidth]{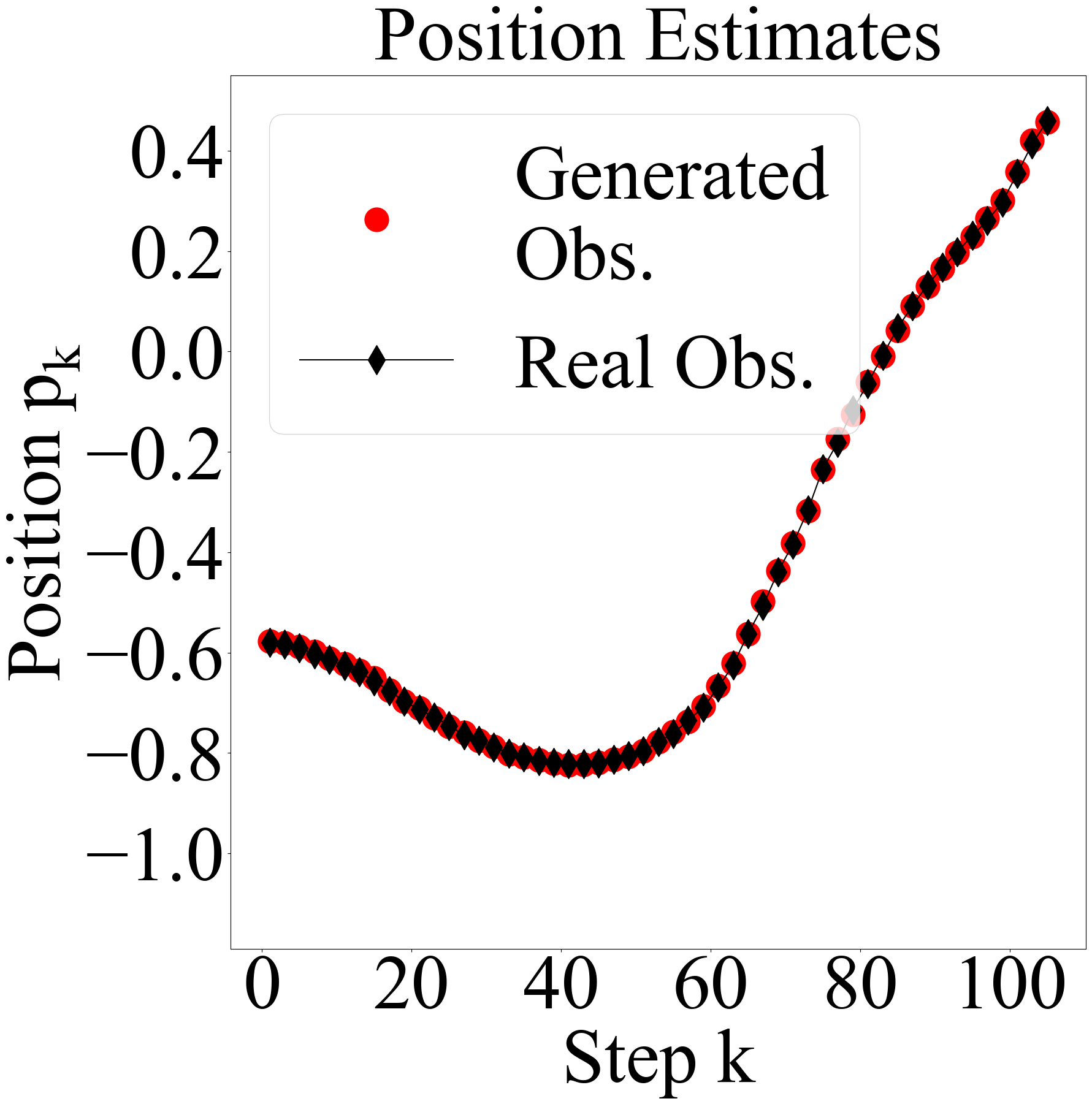}
}
\vspace{-7mm}
}\vspace{-4mm}  
\end{figure}

\subsection{Blur Noise Model}
The blur model, $g_{b}: y_i^c \times \delta_{b,i} \mapsto y_i^{b}$, map a canonical image to a noised image, for a range of blur intensities, $\delta_{b,i} \in [0, 0.3]$. The blur model is $1 \times 50$ and is trained using BCE and weight decay.
\begin{figure}[t]
  \centering
  \floatconts{fig:mc_eval_blur2}
{\caption{Additional evaluation of generative blur models on MC. Figures show estimated position trajectories (using modeled vs. real observations) in simulations. Figure ($a$) shows a trajectory under no blur, ($b$) shows a trajectory under moderate blur, and ($c$) shows a trajectory under high blur. }}
{%
\centering

\subfigure[No blur ($\delta_b=0$)][b]{%
\label{fig:mc_eval_blur2_1}
\captionsetup{justification=centering}
\includegraphics[width=.315\linewidth]{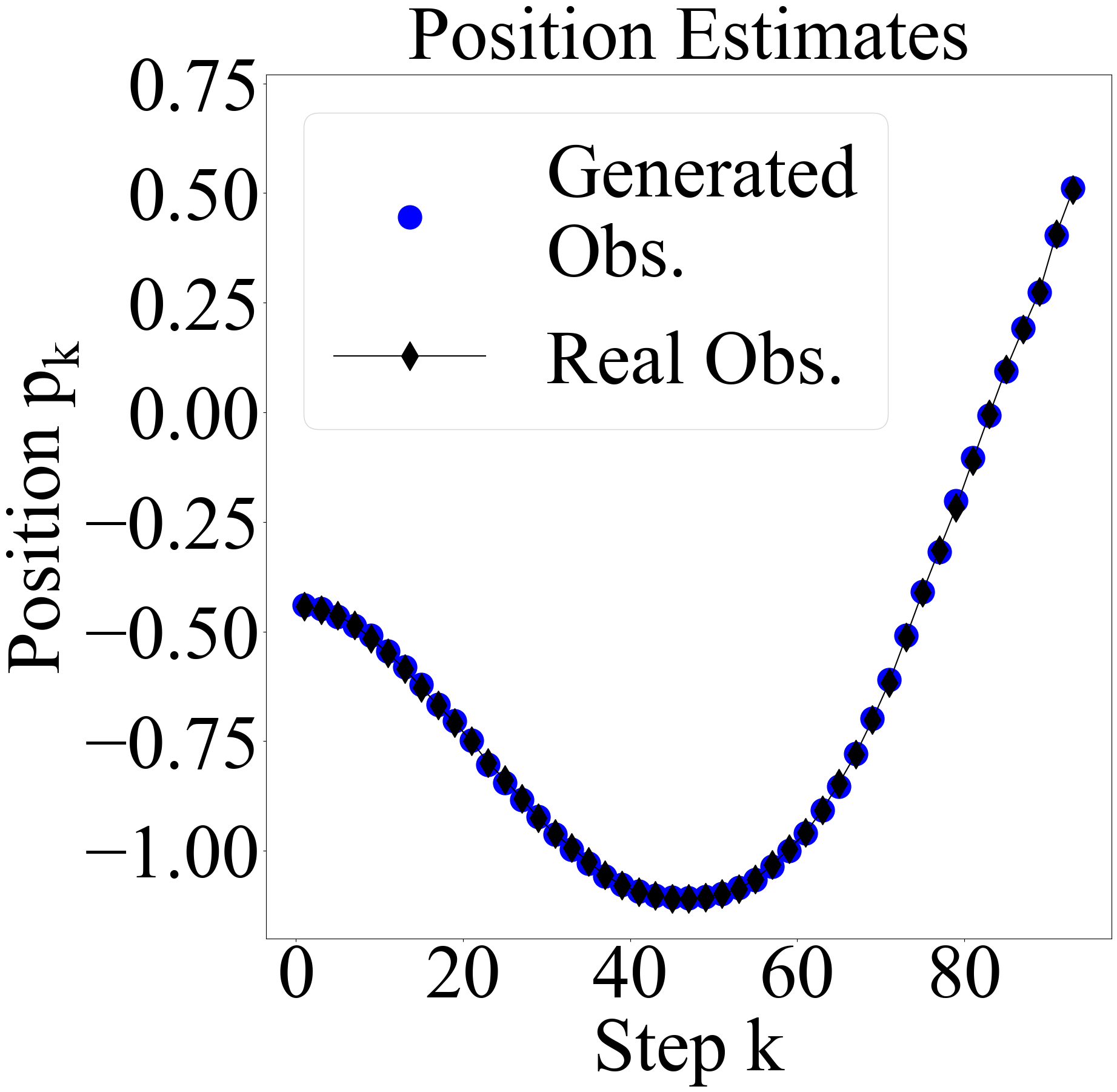}
}
\subfigure[Moderate blur ($\delta_b=0.15$)][b]{%
\label{fig:mc_eval_blur2_2}
\includegraphics[width=.315\linewidth]{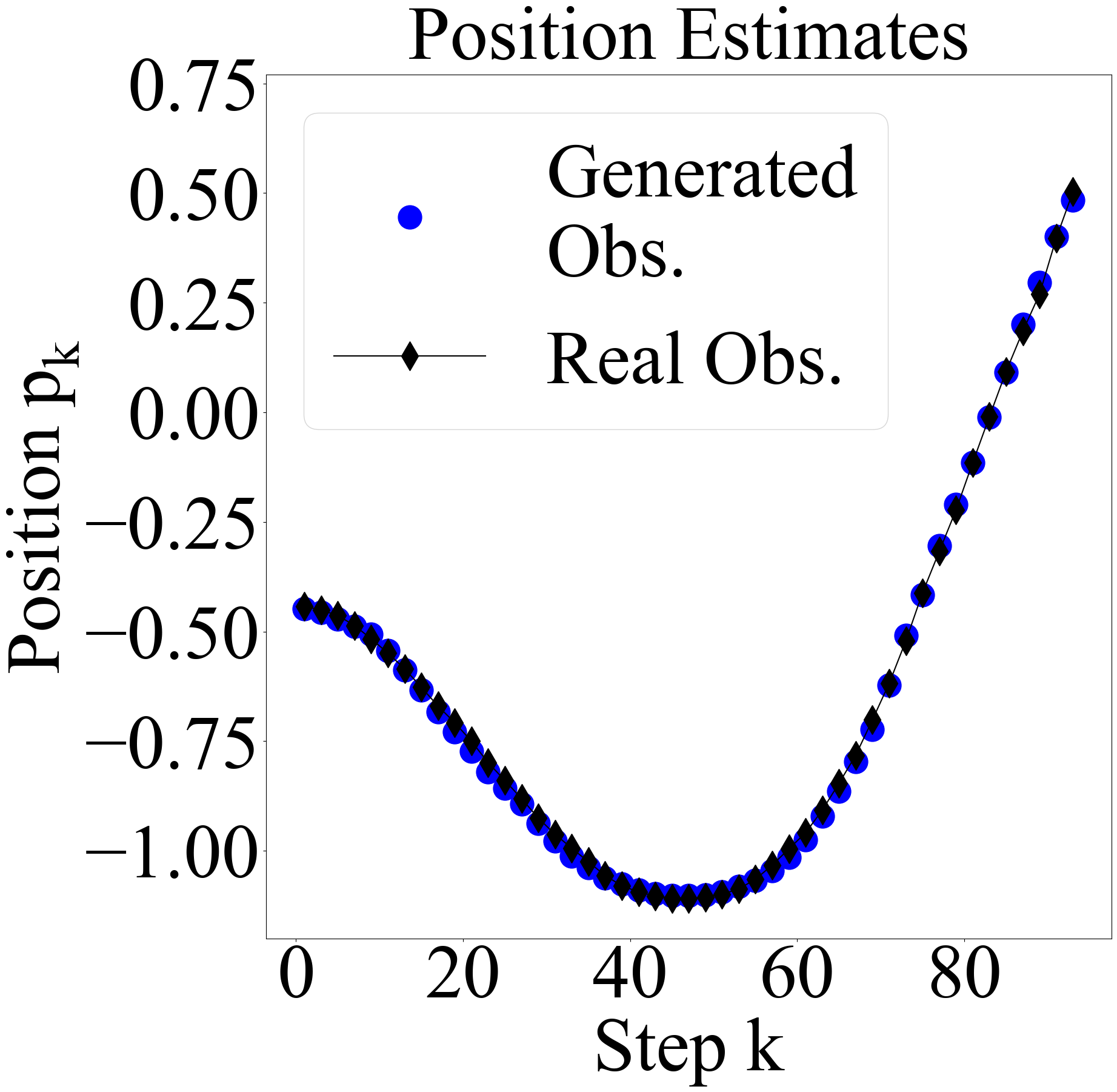}
}
\subfigure[High blur ($\delta_b=0.3$)][b]{%
\label{fig:mc_eval_blur2_3}
\includegraphics[width=.32\linewidth]{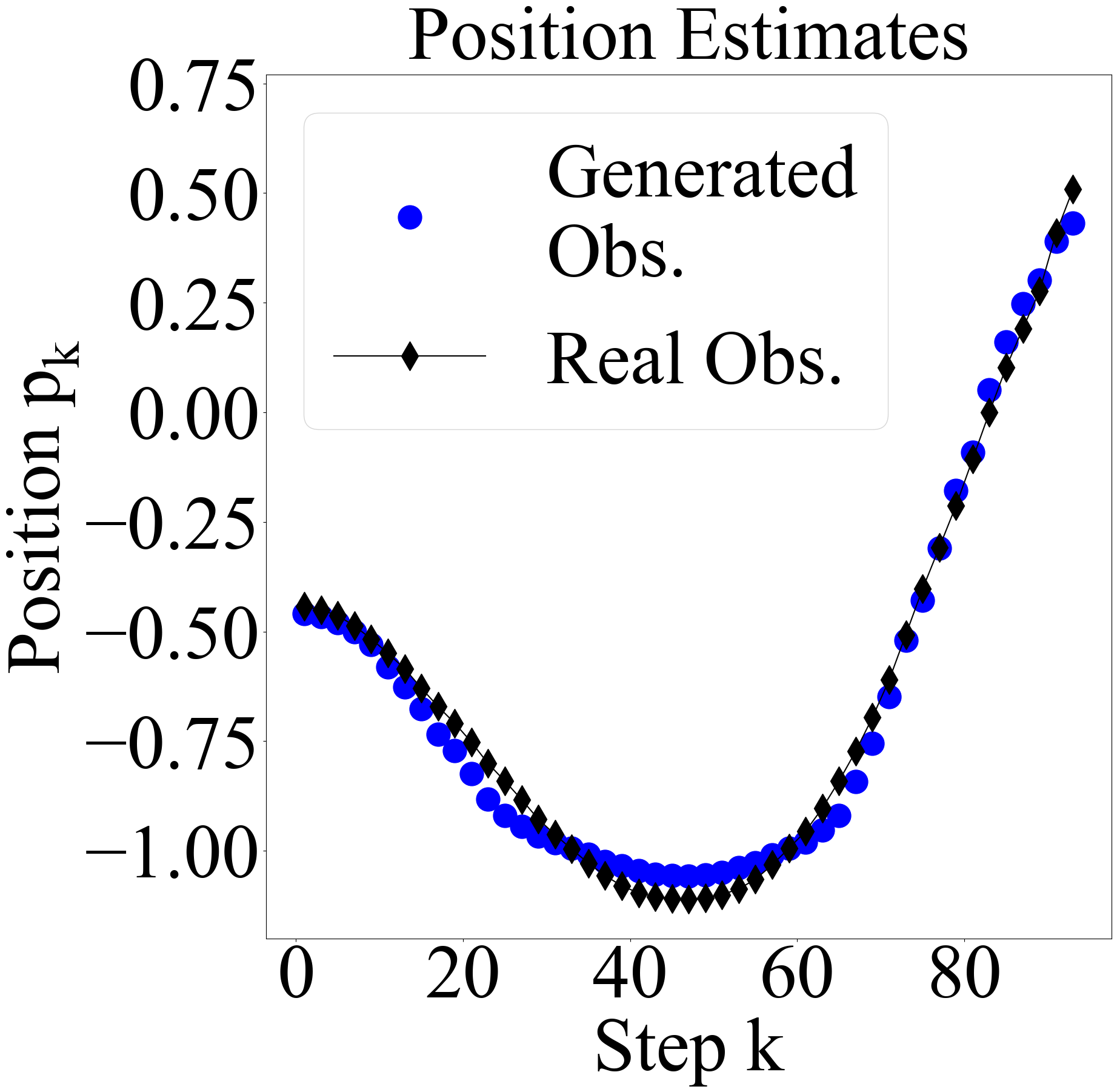}
}
\vspace{-7mm}
}\vspace{-4mm}  
\end{figure}

\subsection{Composite Model}
Interestingly, we do not train a separate composite model but rather compose the contrast and the blur models, $g_{ct}(\delta_{ct}) \circ g_b(\delta_b)$, in order to demonstrate that one can generate complex noise patterns by composing individual noise models. Figures~\ref{fig:mc_composite_generative} provide an evaluation of the composite model in terms of the state-estimation-based metric, on restricted domain for noise parameters (namely, $\delta_c \in[.5, 2]$ and $\delta_b \in [0, 0.15]$). Since the two models were only trained on canonical images, the composition quality deteriorates for larger noise intensities. The figures show the composite model performs similarly well to the single noise models such that composing individual models is a promising approach for modeling complex noise patterns.

\begin{figure}[t]
  \centering
  \floatconts{fig:mc_composite_generative}
{\caption{Evaluation of composite noise model on MC. Figures (a) and (b) show estimated position trajectories (using modeled vs. real observations) in simulations with noise parameters $\delta_{c}=0.5$ and $\delta_{b}=0.15$. Figure (c) shows absolute position estimation errors over 250 simulated trajectories}}
{%
\centering

\subfigure[Trajectory ($p_0=-0.4$)][b]{%
\label{fig:mc_eval_comp_1}
\captionsetup{justification=centering}
\includegraphics[width=.33\linewidth]{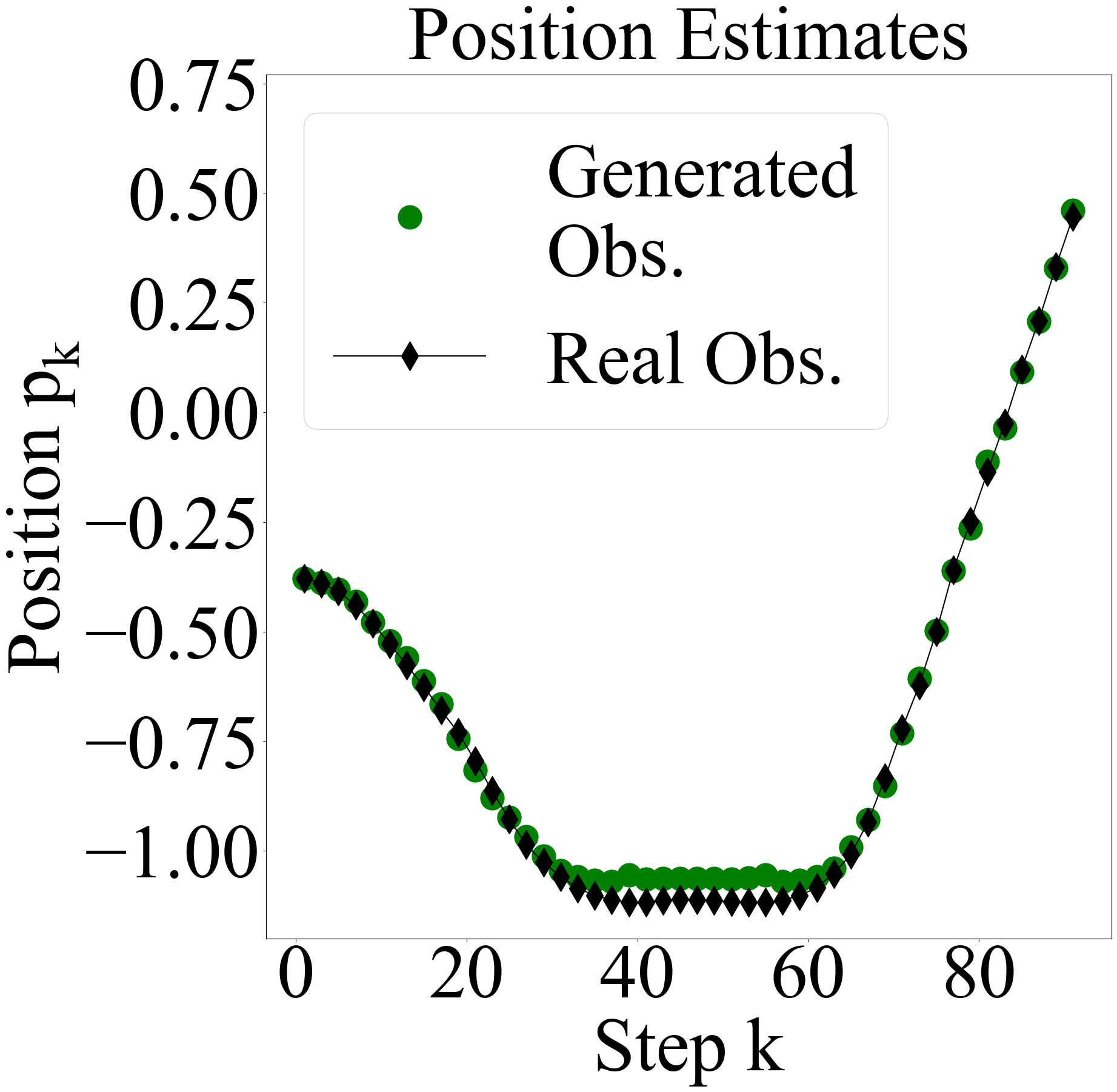}
}
\subfigure[Trajectory ($x_0=-0.59$)][b]{%
\label{fig:mc_eval_comp_2}
\includegraphics[width=.33\linewidth]{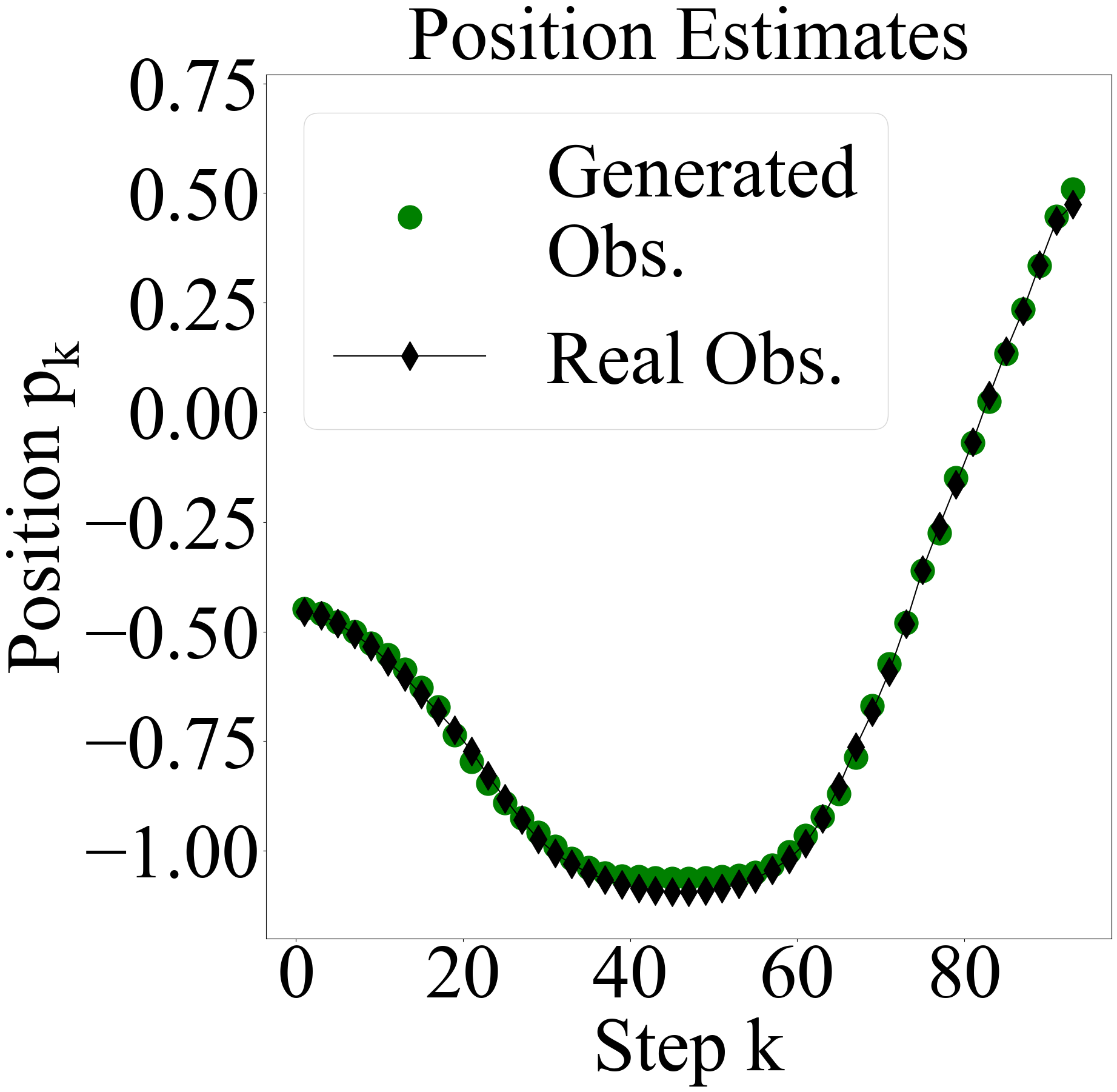}
}
\subfigure[Histogram of estimate errors][b]{%
\label{fig:mc_eval_comp_hist}
\includegraphics[width=.30\linewidth]{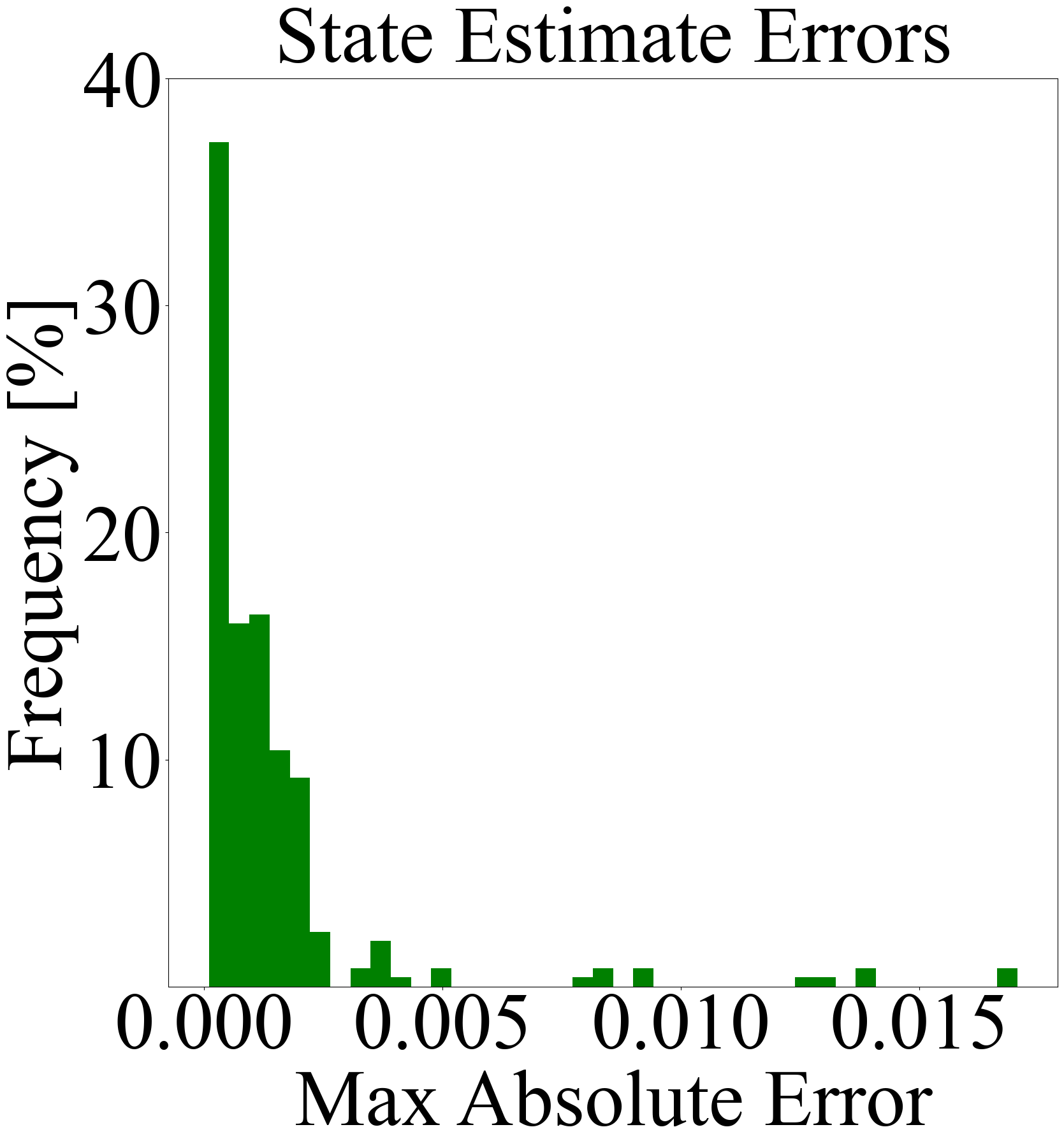}
}
\vspace{-7mm}
}\vspace{-4mm}  
\end{figure}
\subsection{Full System}

The full closed-loop system is described below:
\begin{align}
\begin{split}
p_{k+1} &= p_k + v_{k}\\
v_{k+1} &= v_k + 0.0015u_k - 0.0025*cos(3p_k)\\
y_k &= g(p_k)\\
\hat{p}_k &= h_e(y_k)\\
u_k &= h_c(\hat{p}_k, v_k),
\end{split}
\end{align}
where $g = g_{ct}(\delta_{ct}) \circ g_c$ in the case of contrast noise, $g = g_{b}(\delta_{b}) \circ g_c$ in the case of blur noise, ${g =g_{b}(\delta_b)\circ   g_{ct}(\delta_{ct})  \circ g_c}$ in the case of composite noise.

\subsection{Verification}
We would like to verify that the car reaches the goal with a reward of at least 90. The reward is $r_k = -0.1u_k^2$ for each step before the goal. If the car reaches the goal, $r_k = 100$. Since verifying liveness properties is undecidable in general, we formulate the property as a bounded liveness property, which can be cast as a safety property: $$(p_0 \in [-0.59, -0.4]) \Rightarrow (k = 200 \Rightarrow (p_k \ge p^* \wedge r_k \ge 90)),$$ where $p^* = 0.45$ is the goal position. Note that the 200-step bound on $k$ was found to be conservative and could be changed depending on the controller. Moreover, for the composite noise system, we verify over the range $p_0 \in [-0.55, -0.4]$ given the additional scalability cost of having both contrast and blur networks present in the verification loop and given the controller is less robust for starting positions nearer $-0.59$.

Since computing reachable sets for entire initial range results in too much uncertainty, we split in the initial range into sub-intervals of size $0.00025$ and verify each sub-interval as a separate verification instance. If an instance did not terminate within 24 hours, we split that interval into five sub-intervals and repeated. All instances are verified safe; summary statistics are provided in Table~\ref{tab:verification}. Note the increased verification time for the Blur vs the Contrast models is due to using a cluster-based computing resource for Blur instead of the servers mentioned in the body of the paper.

\section{F1/10 Car}

This section explains the F1/10 case study.

\subsection{Dynamics}
The car dynamics are borrowed from \cite{ivanov20a} and are assumed to follow a bicycle model (with no slipping/friction) as follows:
\begin{align}
  \label{eq:f1_dynamics}
  \begin{split}
    \dot{x_1} &= v cos(\theta)\\
    \dot{x_2} &= v sin(\theta)\\
    \dot{v} &= - c_av + c_ac_m(u - c_h)\\
    \dot{\theta} &= \frac{V}{l_f + l_r}tan(\delta),\\
  \end{split}
\end{align}
where $v$ is linear velocity, $\theta$ is orientation, $x_1$ and $x_2$ are the two-dimensional position; $u = 16$ is the throttle input, and $\delta$ is the heading input (ranging between -15 and 15 degrees); $c_a$ is an acceleration constant, $c_m$ is a car motor constant, $c_h$ is a hysteresis constant, and $l_f$ and $l_r$ are the distances from the car's center of mass to the front and rear, respectively. Parameter values were identified by \cite{ivanov20a} as: ${c_a = 1.633}, {c_m = 0.2}, {c_h = 4}, {l_f = 0.225m}, {l_r = 0.225m}$. Note that the car model is in continuous time; however, the controller is sampled at discrete time steps (at 10Hz), the system effectively operates in discrete time.

\subsection{Canonical LiDAR Model}
The canonical LiDAR model was developed by \cite{ivanov20a} for a square track with width 1.5m and length 20m. The LiDAR model is hybrid since there are four walls that each LiDAR ray could possibly hit:
\begin{align}
  \begin{split}
    y_k^i = g_c(x_k, y_k, \theta_k) = \left\{ \begin{array}{lll} d^r_k/cos(90 + \theta_k + \alpha_i) &&\text{if } \theta_k + \alpha_i \le \theta_r \\
      d^b_k/cos(180 + \theta_k + \alpha_i) &&\text{if } \theta_r < \theta_k + \alpha_i \le -90 \\
      d^t_k/cos(\theta_k + \alpha_i) &&\text{if } -90 < \theta_k + \alpha_i \le \theta_l \\
      d^l_k/cos(90 - \theta_k - \alpha_i) &&\text{if } \theta_l < \theta_k + \alpha_i,\end{array} \right.
  \end{split}
\end{align}
where $d^t_k, d^b_k, d^l_k, d^r_k$ are distances to the four walls; the $\alpha_i$ denote the relative angles for each ray with respect to
the car's heading, i.e., $\alpha_1 = -115, \alpha_2 = -103.5, \dots, \alpha_{21} = 115$; $\theta_l$ and $\theta_r$ are the angles to the left and right corners of the current hallway, respectively. Please consult \cite{ivanov20a} for a full model description. 

\subsection{Controllers}
We use the 12 pre-trained controllers from \cite{ivanov20a}. These controllers were trained using deep deterministic policy gradients (DDPG, \cite{lillicrap15}) and twin delayed DDPG (TD3, \cite{fujimoto18}). The controllers are end-to-end, i.e., each controller $h_c: y_k \mapsto \delta_k$ takes a LiDAR scan and outputs a heading. All controllers were verified by \cite{ivanov20a} to be safe for the canonical LiDAR model, yet all controllers resulted in at least some crashes when used on the real platform.

\subsection{Training Set}
The training set consists of 115 real-data trajectories (roughly 10 trajectories per each of the 12 controllers). The original trajectories contain only (full 1081-dimensional) LiDAR scans. Since the proposed data-driven modeling approach requires car states as well, we used a particle filter (as presented by \cite{thrun05}) to estimate the car poses. Given the state estimates, the final labeled dataset is ${\mathcal{D} = \{(x_{1,i}, x_{2,i}, \theta_i, y_i)\}}$, where $(x_{1,i}, x_{2,i})$ denote the car's $(x,y)$-coordinates and $\theta_i$ is the car's orientation.

\subsection{Adversarial LiDAR Model}
To train the LiDAR noise model, we label each LiDAR scan according to the procedure in Section \ref{sec:modeling}. Given a canonical scan $g_c(x_{1,i}, x_{2,i}, \theta_i)$, we set ${n_i^d = 1}$ if ${|g_c(x_{1,i}, x_{2,i}, \theta_i)^d - y_i^d| > t_{\Delta}^*}$, and ${n_i^d = 0}$, otherwise. The threshold $t_{\Delta}^*$ is chosen to minimize the error between noised canonical scans and real scans over the training data. Using the labeled training set ${D_l = \{((g_c(x_{1,i}, x_{2,i}, \theta_i),x_i), n_i)\}}$, we train classifiers of increasing sizes (listed in Table~\ref{tab:f1_train_crash_all}) using weighted BCE (to account for class imbalance between adversarial and non-adversarial examples). The adversarial example (class 1) loss is weighted 4 times the class 0 loss.

\begin{wraptable}{r}{0.55\textwidth}
\vspace{0px}
\scriptsize
\centering
\begin{tabular}{|r|c|c|c|c|c|}
    \hline
        \multicolumn{6}{|c|}{\textbf{Noiser Training Results}}\\
            \hline
        Model Size & 3x100 & 4x100 & 5x100  &  6x100 & 6x200\\
        \hline
        WSA [\%] & 61& 77 & 87 & 98 & 100 \\
        \hline
        Control Err. & 3700 & 2312 & 1485 & 774 &  594 \\
        \hline
        \multicolumn{6}{|c|}{\textbf{Safe Trajectory Distribution [\%]}} \\
        \hline
        Controller:\textbf{ Real Data} & \multicolumn{5}{|c|}{\textbf{Simulated Data Using Noiser Model}} \\
        \hline
        DDPG, 64x64, 1:  \textbf{0} & 19 & 25 & 16 & 15 &  8 \\
        \hline
        DDPG, 64x64, 2: \textbf{20}  & 31 & 43 & 31 & 25 &  20 \\
        \hline
        DDPG, 64x64, 3: \textbf{80}  & 88 & 83 & 88 & 82 &  97 \\
        \hline
        DDPG, 128x128, 1: \textbf{80}  & 75 & 67 & 69 & 62 &  63 \\
        \hline
        DDPG, 128x128, 2:  \textbf{40} & 89 & 84 & 88 & 85 &  88 \\
        \hline
        DDPG, 128x128, 3:  \textbf{0} & 29 & 29 & 29 & 32 &  27 \\
        \hline
        TD3, 64x64, 1:  \textbf{90} & 78 & 79 & 64 & 82 &  100\\
        \hline
        TD3, 64x64, 2:  \textbf{90} & 75 & 80 & 92 & 77 &  74 \\
        \hline
        TD3, 64x64, 3:  \textbf{90} & 94 & 94 & 97 & 72 &  100 \\
        \hline
        TD3, 128x128, 1:  \textbf{90} & 100 & 100 & 95 & 96 &  100 \\
        \hline
        TD3, 128x128, 2:  \textbf{90} & 85 & 78 & 98 & 91 &  91 \\
        \hline
        TD3, 128x128, 3:  \textbf{90} & 87 & 91 & 95 & 86 &  100 \\
        \hline
    \end{tabular}
    \vspace{-5px}
    \caption{Training results and crash distributions for different noise models. Pre-trained controllers and real data were borrowed from \cite{ivanov20a}.}
    \label{tab:f1_train_crash_all}
    \vspace{-10px}
\end{wraptable}

Figures~\ref{fig:f1_model_eval_control_all} show an evaluation of the adversarial model used the control-based metric, over most controllers.  The figures illustrate that the classifier-based adversarial model captures most of the control-action variability due to the adversarial noise (especially when compared with the canonical model).

As further evaluation, Figures~\ref{fig:f1_model_eval_trajectory_all} also show the evaluation using the trajectory-based metric. Notice that overall the trajectories look quite similar; in particular most crashes observed in the real world correspond to crashes observed in simulation, too.

\subsection{Denoiser}
The denoiser's goal is to take a noised LiDAR scan and convert it to a canonical scan. We train a $1\times 100$ fully-connected denoiser NN, $h_d: y_i \mapsto \hat{y}_i$, on pairs $(g_c(x_{1,i}, x_{2,i}, \theta_i), y_i)$ using BCE loss and weight decay.

\subsection{Full System}
The full system is described below:
\begin{align}
\begin{split}
x_{1,k+1}, x_{2,k+1}, v_{k+1}, \theta_{k+1} &= f(x_k, y_k, v_k, \theta_k, \delta_k)\\
y_k &= g_c(x_{1,k}, x_{2,k}, \theta_{k})\\
y_k^a &= g_n(x_{1,k}, x_{2,k}, \theta_{k}, y_k)\\
\hat{y}_k &= h_d(y_k^a)\\
\delta_k &= h_c(\hat{y}_k),
\end{split}
\end{align}
where $f$ is obtained by following the continuous-times dynamics in~\eqref{eq:f1_dynamics} for 0.1s.

\subsection{Verification}
We verify that the car can navigate the first right-hand turn. In particular, the car is started in a .2 m interval in the middle of the first segment (centered at the origin), 9m from the front wall ($x_{2,0}=1$). We verify that the car reaches 2.5m from the left wall in the second hallway (since that is roughly how the where the real data ends, i.e., the generative model is not reliable beyond that point). Similar to the MC case, we formulate the liveness property as a bounded safety property: $$(x_{1,0} \in [-0.1, 0.1]) \Rightarrow (\phi(x_{1,k}, x_{2,k}) \wedge (k = 70 \Rightarrow x_{1,k} \ge 1.75)),$$ where $\phi(x_{1,k}, x_{2,k})$ is the safety property of no crashes. 

Verifying the F1/10 car safety is particularly challenging both due the complex LiDAR model and due to the adversarial noise classifier. All of these components introduce possible branching in model: if the uncertainty is too high, a LiDAR ray could reach different walls, in which case both models need to be verified separately. Similar to the case of MC, we split the initial range into sub-intervals of size $0.000025$ and sub-divide each interval if verification takes longer than 24 hours. The minimum interval sized used was $0.000001$.  Summary statistics are provided in Table~\ref{tab:verification}.
\begin{figure}[t]
  \centering
   
  \floatconts{fig:f1_model_eval_control_all}
{\caption{Control-based evaluation of F1/10 noise model, using 10 controllers from \cite{ivanov20a}. Figures outputs from each controller on a training trajectory for canonical, noised and real measurements. }\label{fig:f1_simvreal}}
{%
\centering
\subfigure[DDPG, 64, 1][b]{%
\label{fig:f1_simvreal1}
\includegraphics[width=.44\linewidth]{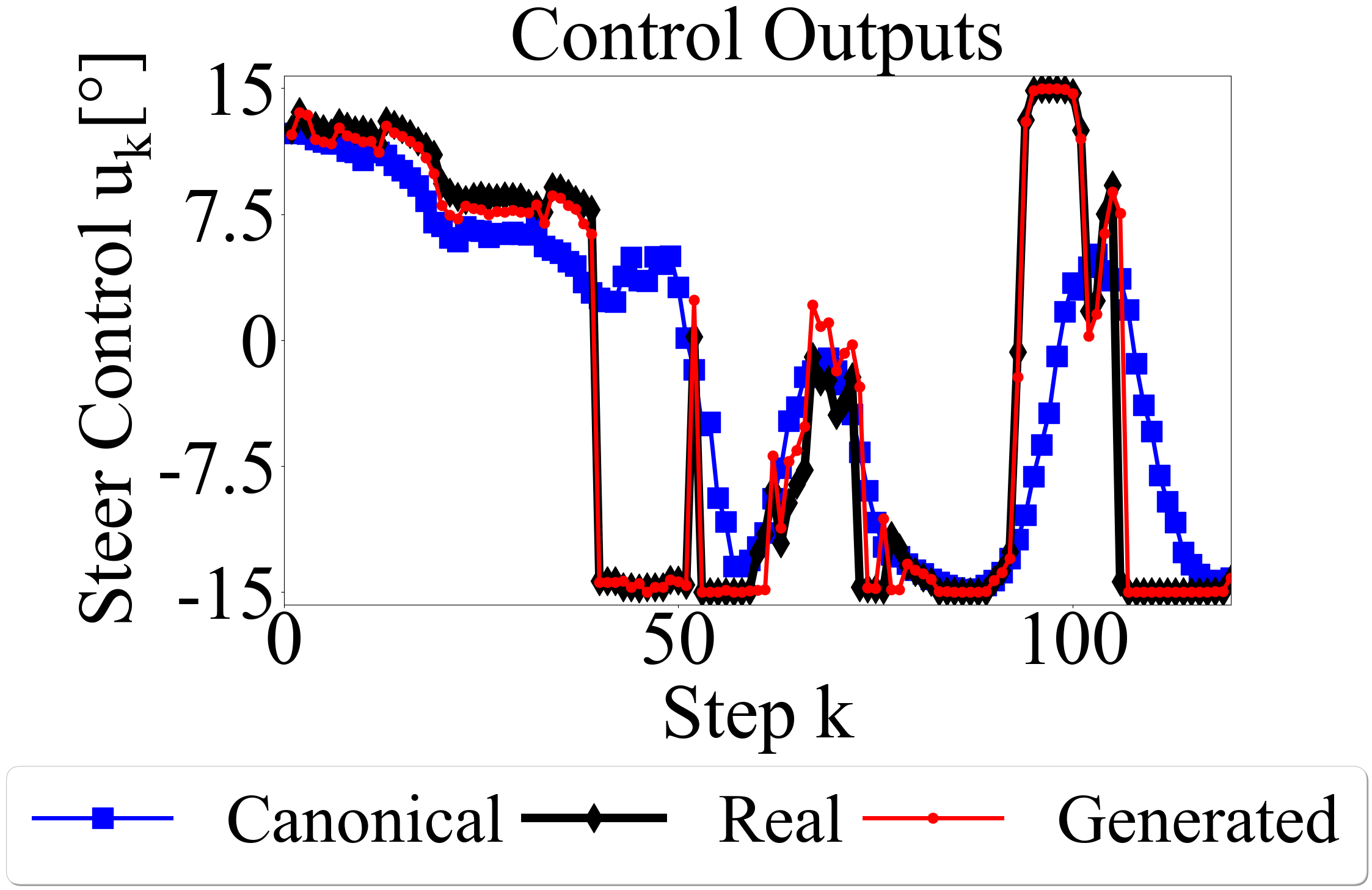}
}
\hfill
\subfigure[DDPG, 64, 2][b]{%
\label{fig:f1_simvreal3}
\includegraphics[width=.44\linewidth]{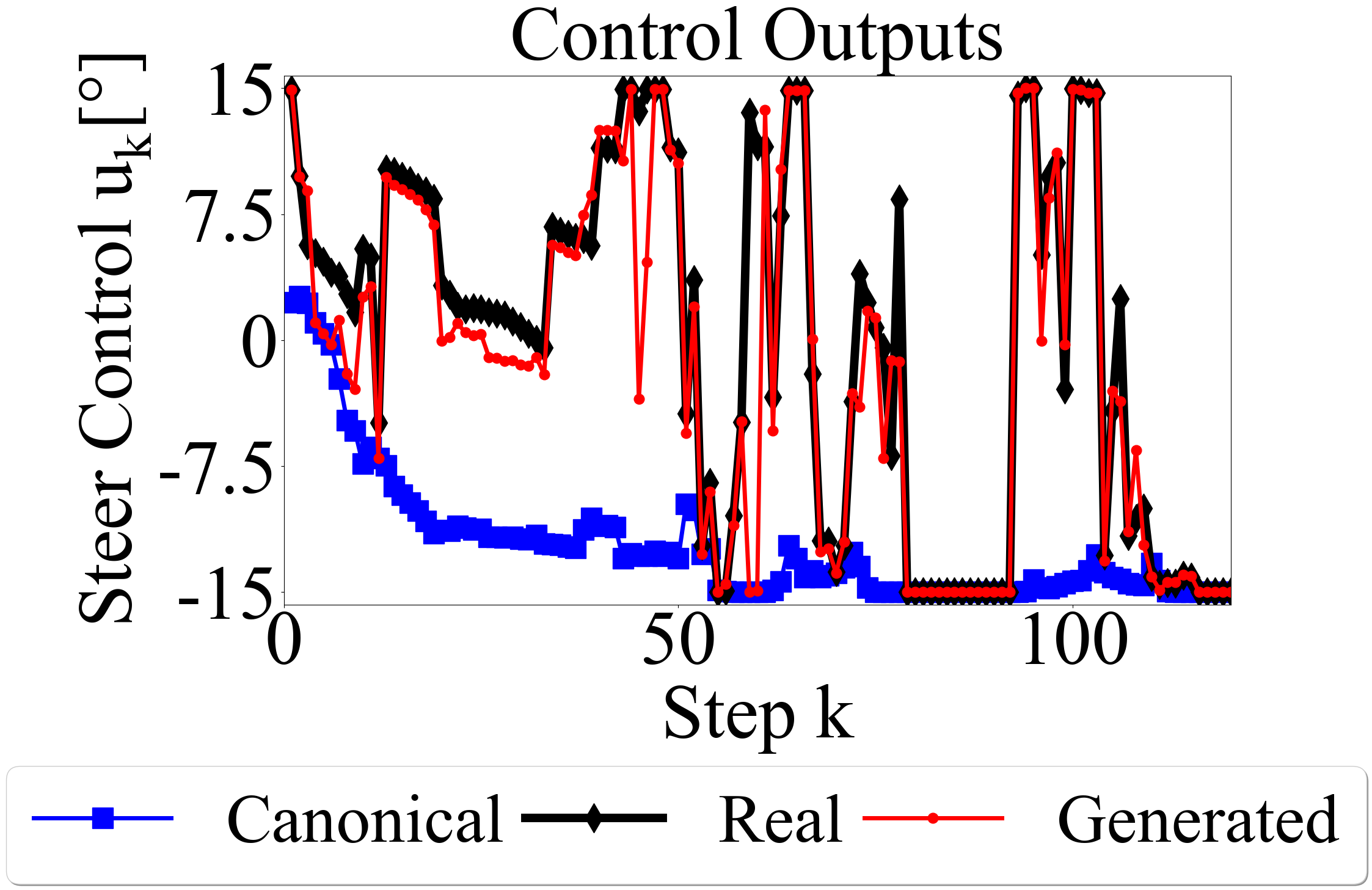}
}
\hfill
\subfigure[DDPG, 64, 3][b]{%
\label{fig:f1_simvreal4}
\includegraphics[width=.44\linewidth]{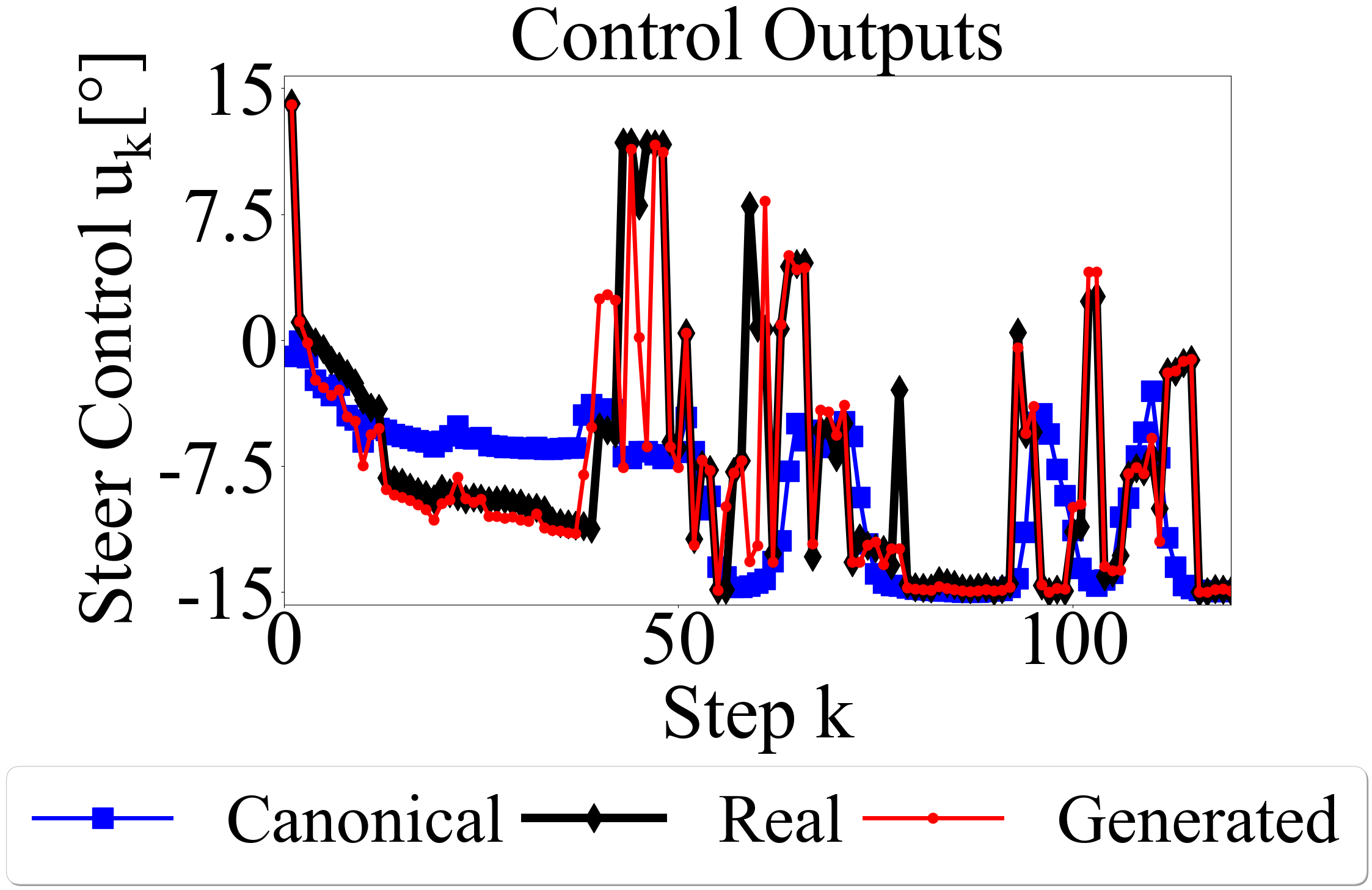}
}
\hfill
\subfigure[DDPG, 128, 1][b]{%
\label{fig:f1_simvreal2}
\includegraphics[width=.44\linewidth]{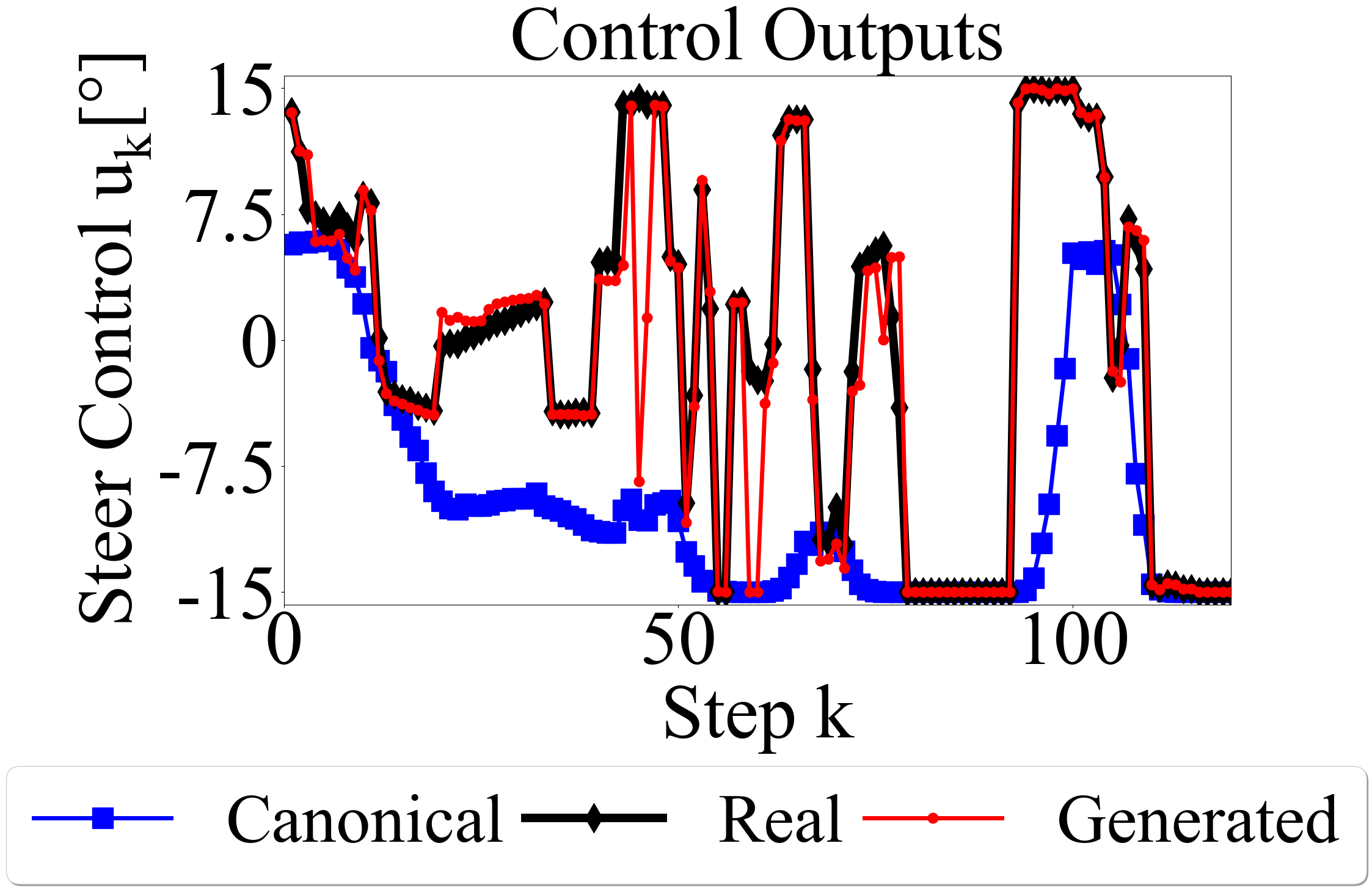}
}
\subfigure[DDPG, 128, 2][b]{%
\label{fig:f1_simvreal1}
\includegraphics[width=.44\linewidth]{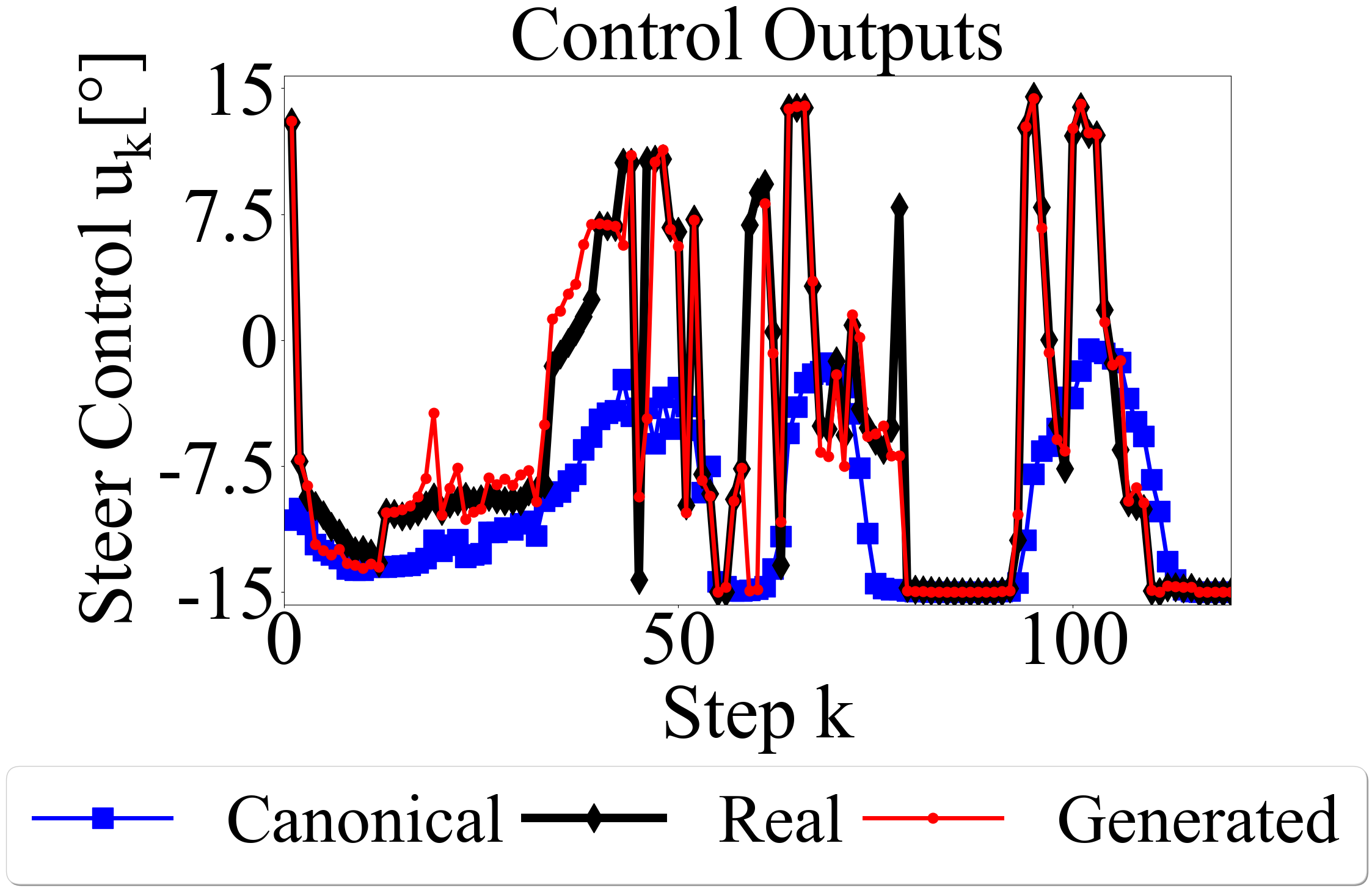}
}
\hfill
\subfigure[DDPG, 128, 3][b]{%
\label{fig:f1_simvreal3}
\includegraphics[width=.44\linewidth]{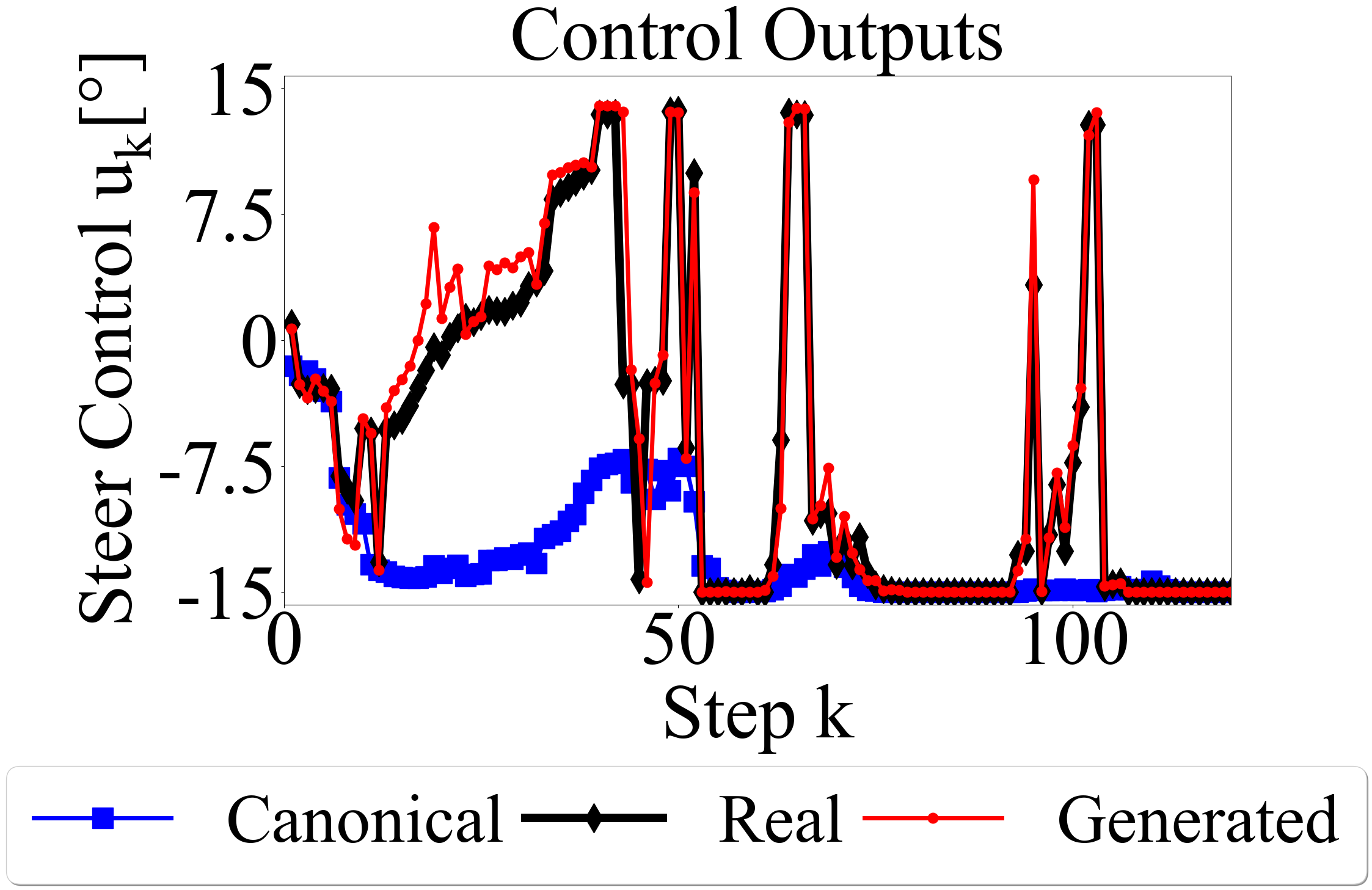}
}
\hfill
\subfigure[TD3, 64, 1][b]{%
\label{fig:f1_simvreal4}
\includegraphics[width=.44\linewidth]{figs/f1_model_eval_traj_TD3_L21_64x64_C1.png}
}
\hfill
\subfigure[TD3, 64, 2][b]{%
\label{fig:f1_simvreal2}
\includegraphics[width=.44\linewidth]{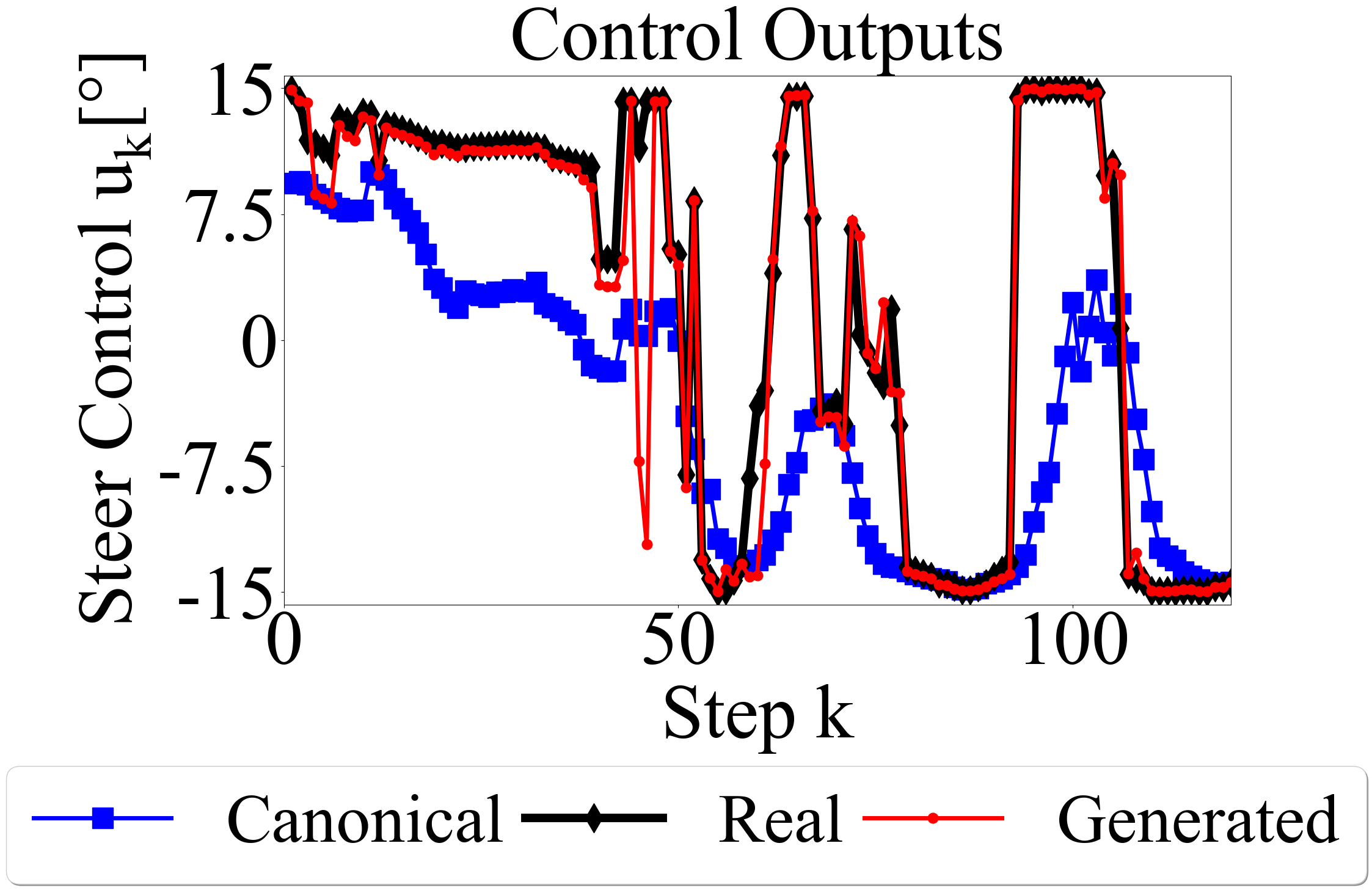}
}
\vspace{-7mm}
}\vspace{-10mm}  
\end{figure}

\begin{figure}[t]
  \centering
   
  \floatconts{fig:f1_model_eval_trajectory_all}
{\caption{Simulated trajectories under the 5x100 noiser model vs. real trajectories for all twelve controllers from \cite{ivanov20a} (without using a denoiser). Bold trajectories indicate those that ended in a crash.}\label{fig:f1_simvreal}}
{%
\centering
\subfigure[DDPG, 64, 1][b]{%
\label{fig:f1_simvreal1}
\includegraphics[width=.23\linewidth]{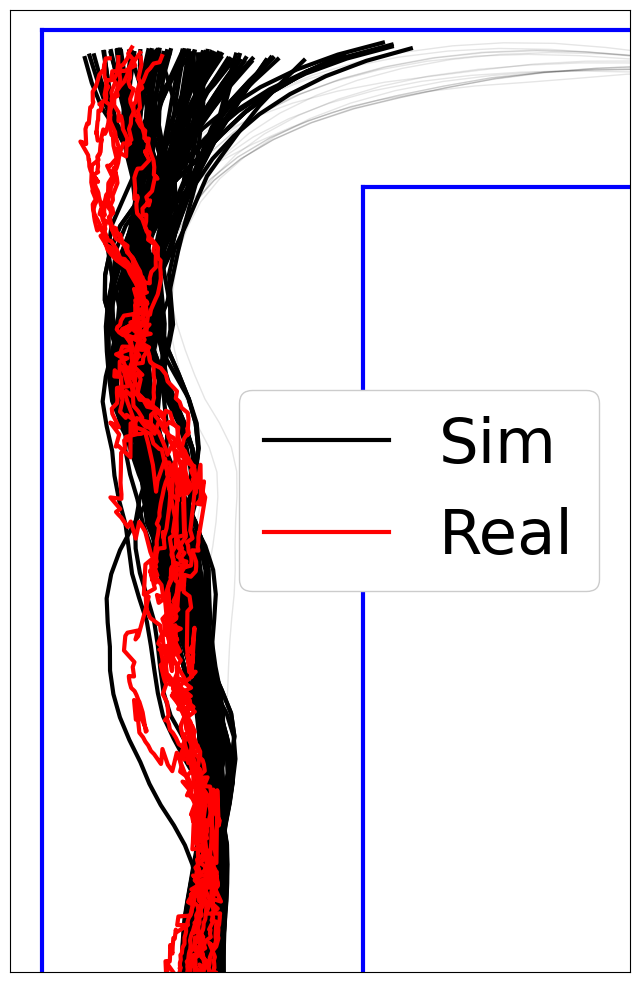}
}
\hfill
\subfigure[DDPG, 64, 2][b]{%
\label{fig:f1_simvreal3}
\includegraphics[width=.23\linewidth]{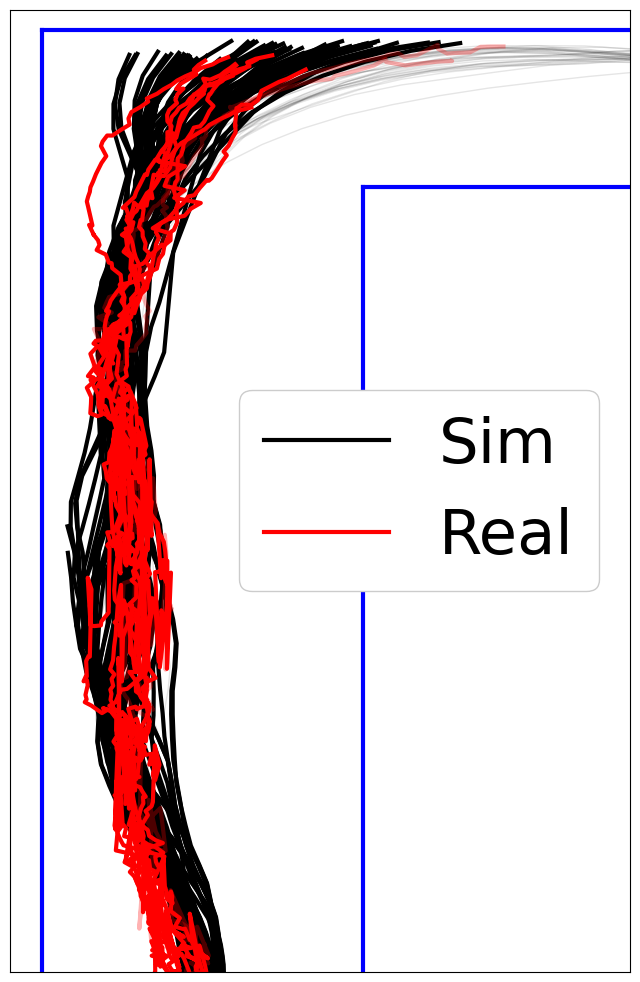}
}
\hfill
\subfigure[DDPG, 64, 3][b]{%
\label{fig:f1_simvreal4}
\includegraphics[width=.23\linewidth]{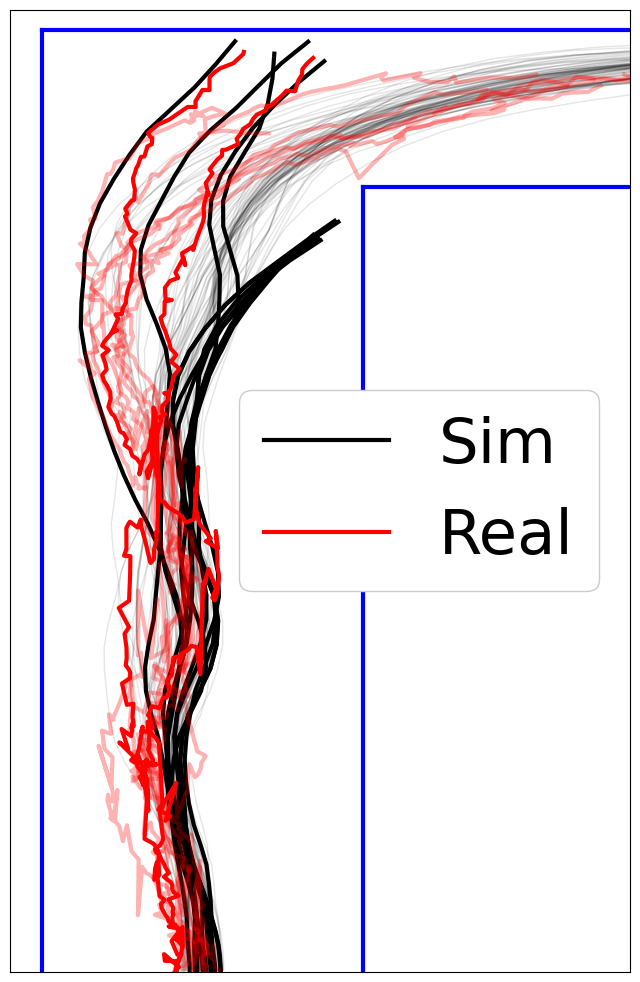}
}
\hfill
\subfigure[DDPG, 128, 1][b]{%
\label{fig:f1_simvreal2}
\includegraphics[width=.23\linewidth]{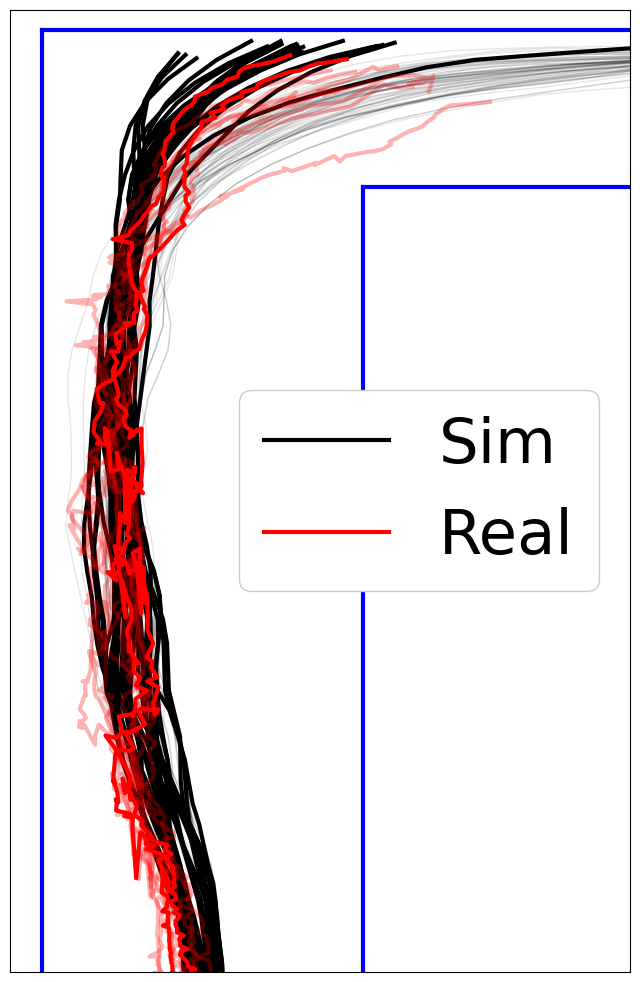}
}
\subfigure[DDPG, 128, 2][b]{%
\label{fig:f1_simvreal1}
\includegraphics[width=.23\linewidth]{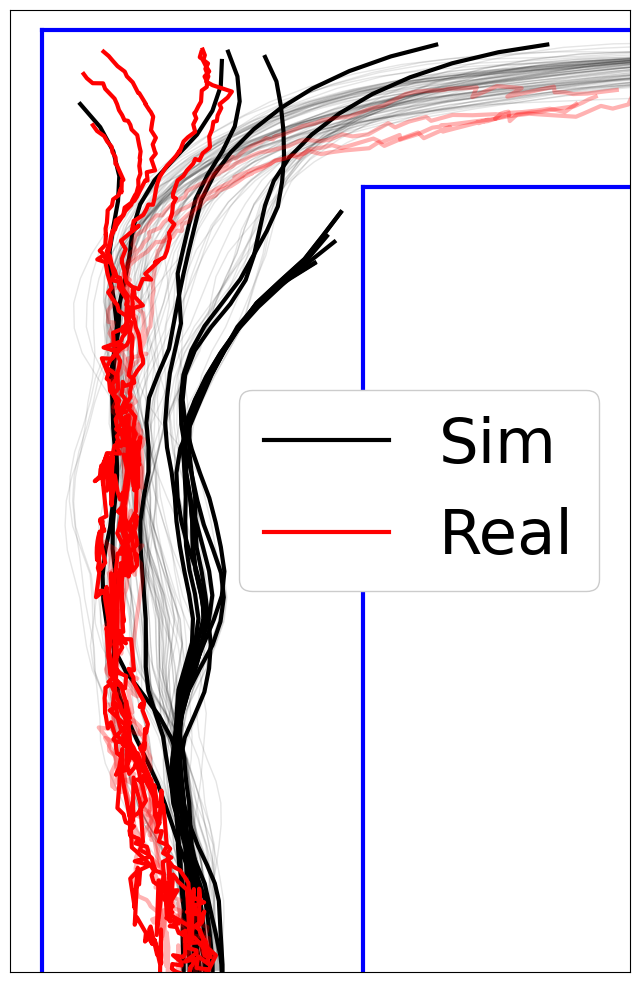}
}
\hfill
\subfigure[DDPG, 128, 3][b]{%
\label{fig:f1_simvreal3}
\includegraphics[width=.23\linewidth]{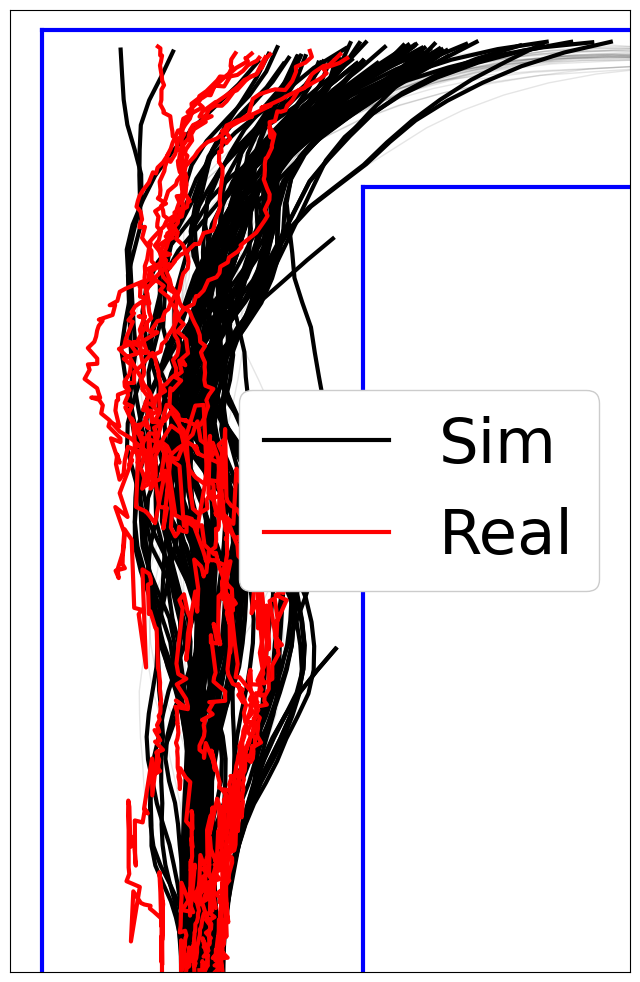}
}
\hfill
\subfigure[TD3, 64, 1][b]{%
\label{fig:f1_simvreal4}
\includegraphics[width=.23\linewidth]{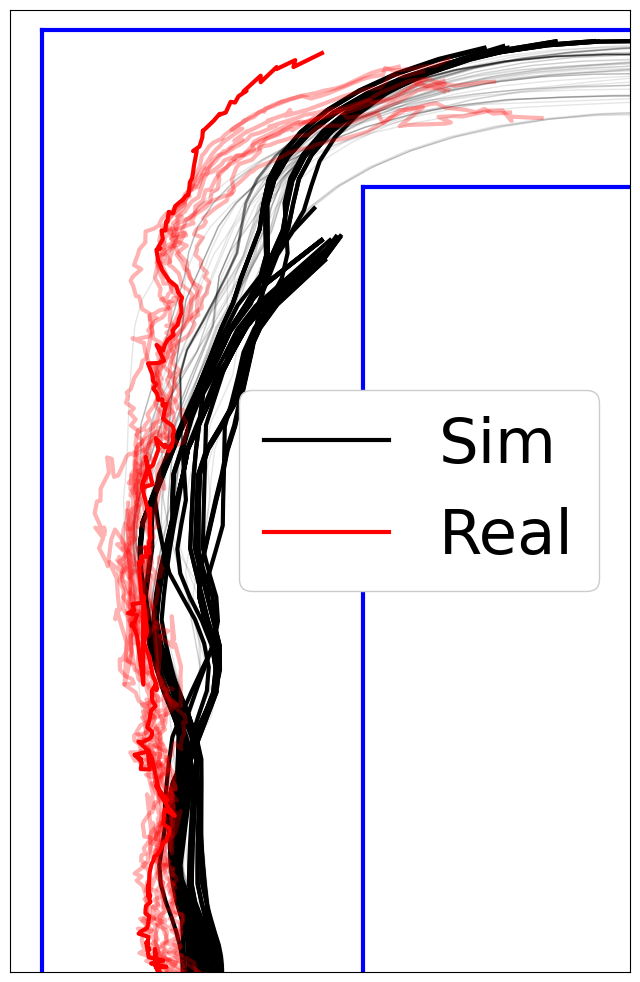}
}
\hfill
\subfigure[TD3, 64, 2][b]{%
\label{fig:f1_simvreal2}
\includegraphics[width=.23\linewidth]{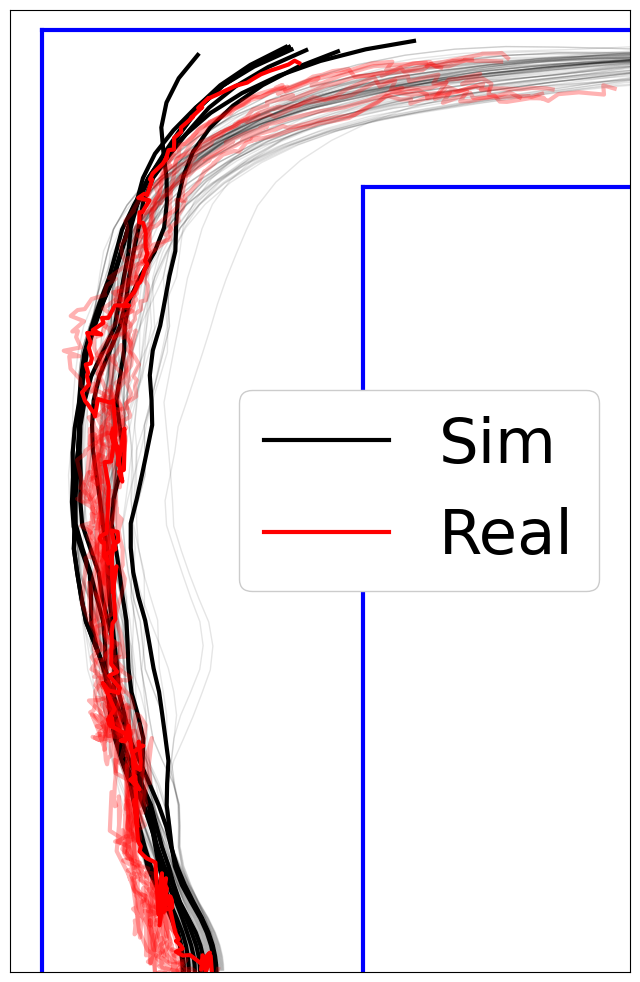}
}
\subfigure[TD3, 64, 3][b]{%
\label{fig:f1_simvreal1}
\includegraphics[width=.23\linewidth]{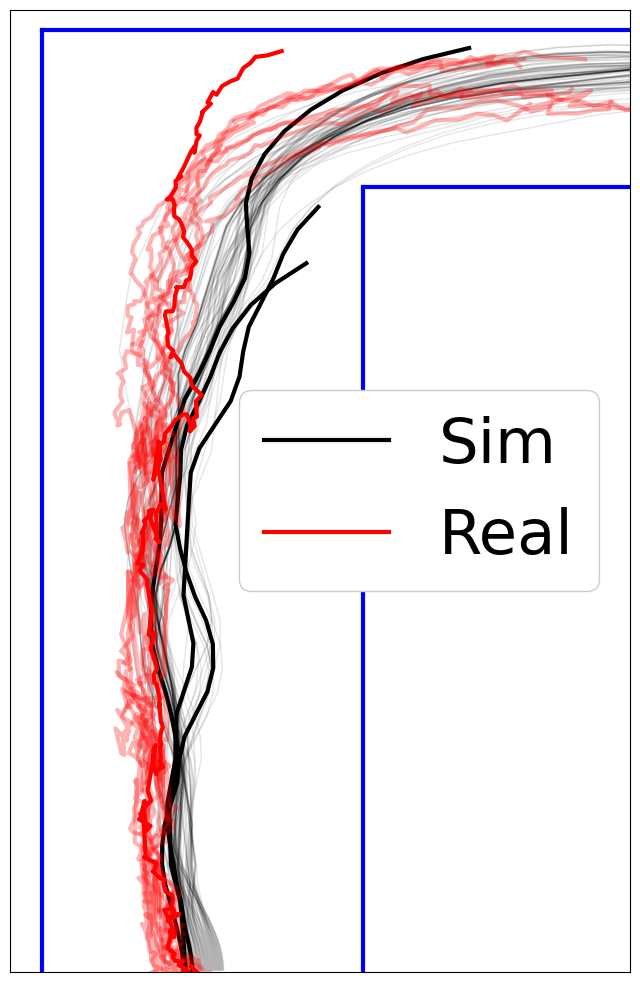}
}
\hfill
\subfigure[TD3, 128, 1][b]{%
\label{fig:f1_simvreal3}
\includegraphics[width=.23\linewidth]{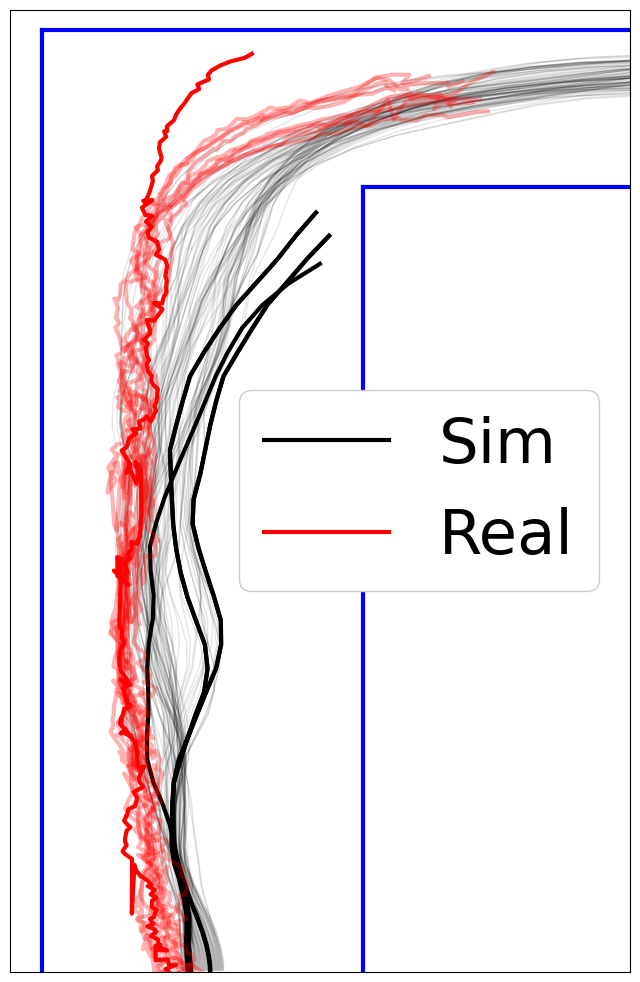}
}
\hfill
\subfigure[TD3, 128, 3][b]{%
\label{fig:f1_simvreal4}
\includegraphics[width=.23\linewidth]{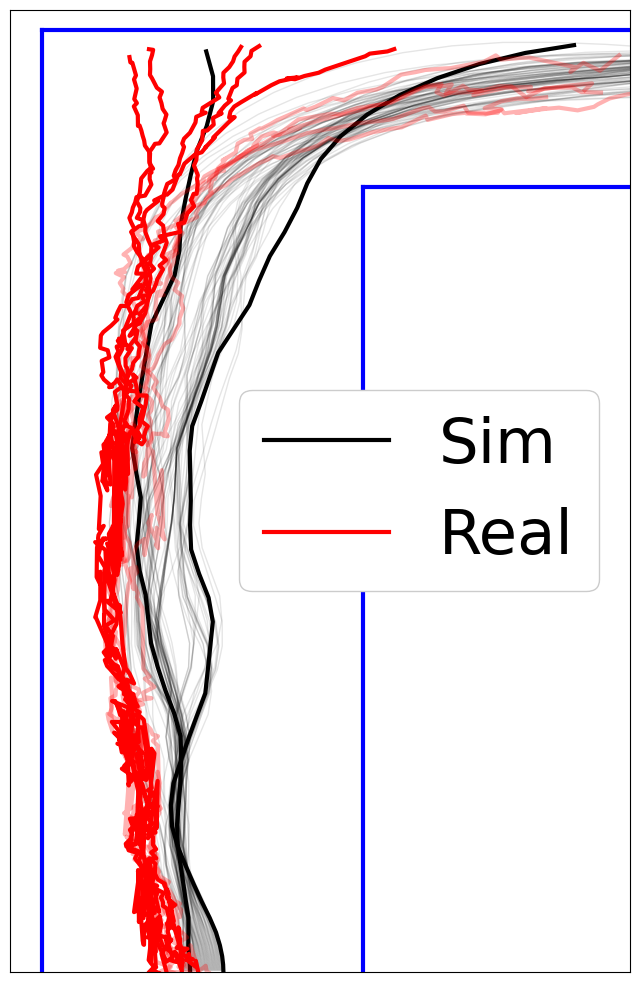}
}
\hfill
\subfigure[TD3, 128, 3][b]{%
\label{fig:f1_simvreal2}
\includegraphics[width=.23\linewidth]{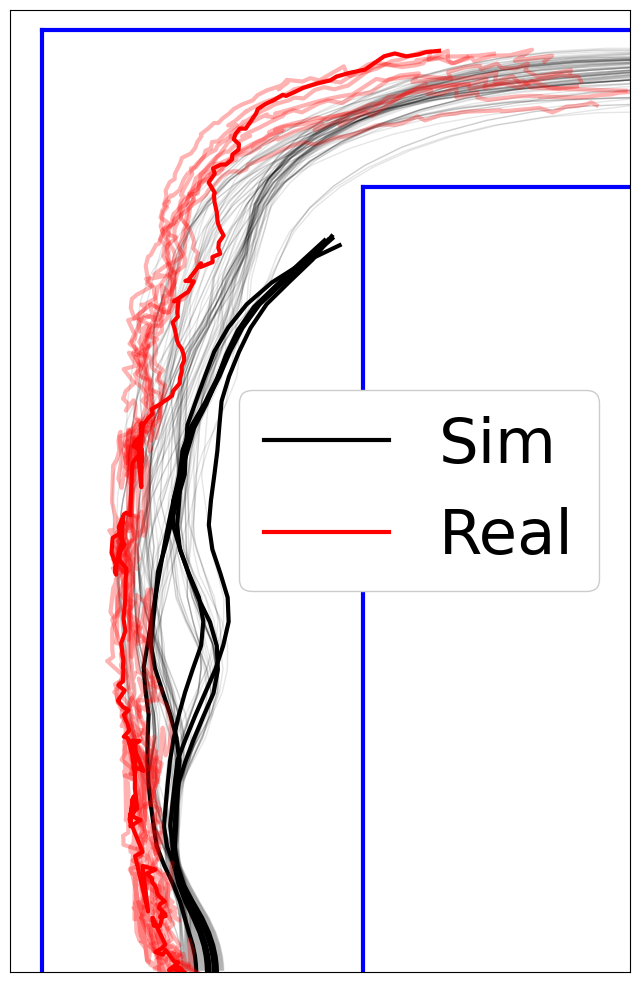}
}
\vspace{-7mm}
}\vspace{0mm}  
\end{figure}

\end{document}